# ixFLIM: Interferometric Excitation Fluorescence Lifetime Imaging Microscopy

Pavel Malý[1*], Dita Strachotová[1], Aleš Holoubek[2], Petr Heřman[1]

[1]*Faculty of Mathematics and Physics, Institute of Physics, Charles University, Prague, Czech Republic*

[2]*Department of Proteomics, Institute of Hematology and Blood Transfusion, Prague, Czech Republic*

*pavel.maly@mff.cuni.cz

**Abstract:** Fluorescence lifetime imaging microscopy (FLIM) is a well-established technique with numerous imaging applications. Yet, one of the limitations of FLIM is that it provides information about the emitting state only. Here, we present an extension of FLIM by interferometric measurement of fluorescence excitation spectra. Interferometric Excitation Fluorescence Lifetime Imaging Microscopy (ixFLIM) reports on the correlation of the excitation spectra and emission lifetime, providing the correlation between the ground-state absorption and excited-state emission. As such, it extends the applicability of FLIM and removes some of its limitations. We introduce ixFLIM on progressively more complex systems and apply it to quantitative resonance energy transfer imaging from a single measurement.

**Introduction**

Fluorescence lifetime imaging microscopy (FLIM) is a sensitive imaging technique well-established in life sciences[1–5]. Usually independent of the fluorophore concentration, it allows *in-vivo* cellular imaging based on intrinsic fluorescence[6] or extrinsic fluorophores[7], probing local environment (such as viscosity, polarity, ion or metabolite concentrations, oxygen imaging, etc.)[8,9], and various intracellular interactions (protein, lipid, DNA, etc.)[10]. A powerful application of FLIM is its use for detection of Förster resonance energy transfer (FRET) by the shortening of the excited state lifetime of the donor fluorophore by the excitation transfer to the acceptor[1,11,12]. Despite its popularity, FLIM has its limitations. An obvious one is the presence of several emitting species with overlapping emission and similar lifetime, which are difficult if not impossible to disentangle in the single time dimension. A serious limitation in the FRET application is the case of highly efficient excitation transfer, which shortens the donor lifetime beyond reliable measurement. Furthermore, a common way to prove the FRET presence is to photo-bleach the acceptor and observe recovery of the longer donor lifetime, with possible photoconversion of the bleached acceptor to a different emitting species being an unwanted, unknown contribution[13].

Variants of FLIM have been and are being developed to address these issues. A straightforward one is FLIM with spectral resolution of the emission that can be realized, e.g., by multichannel detection[14] or by Fourier spectrometry[15]. While useful for the identification of the emitting species, this approach still provides information about the emitting state only, and thus most of the substantial FLIM problems remain. To address the absorption, multiplexing techniques have been developed that combine several excitation wavelengths either electronically[16] or interferometrically[17,18]. These, however, only resolve the excitation at a few fixed wavelengths. Meanwhile, in ultrafast nonlinear spectroscopy an approach to probe excitation at all wavelengths at once while keeping the time resolution is known, using double-pulse excitation as in Fourier Transform spectroscopy[19,20]. Recently, such approach was even applied to single molecules,[21,22] with an example combined with time-correlated single-photon counting (TCSPC)[23].

Here, we present an extension of FLIM with broadband interferometric two-pulse excitation that allows spectrally resolved excitation with all relevant wavelengths present simultaneously within a single excitation pulse. Interferometric Excitation Fluorescence Lifetime Imaging Microscopy (ixFLIM) correlates the excitation spectrum with fluorescence decay within a single measurement. At each pixel (or voxel) of the image, it thus provides an additional dimension along which the individual fluorescent species can be resolved. While ixFLIM keeps the advantages of FLIM, the correlation of excitation spectrum and emission lifetime removes some of its limitations. For sensing and imaging applications, formation of aggregates can be recognized and eventually excluded, and the emitting species can be identified and disentangled by the combination of their lifetime and excitation spectrum. As we demonstrate in detail, ixFLIM is extremely useful in FRET measurements. In FRET-ixFLIM, the information about transfer efficiency is present in two forms: as a rise of the acceptor signal after donor excitation, and as donor excitation spectrum detected by the acceptor emission. In contrast to the common FLIM, this interlinked information allows the experimenter to measure both highly efficient and highly inefficient energy transfer using the same donor-acceptor pair.

**Results**

*Principle of ixFLIM*

The principle of the ixFLIM experiment is outlined in Fig. 1. The imaging and detection part is identical to the standard time-domain FLIM, in our case TCSPC detection is coupled to scanning confocal microscope (Fig. 1a). The spatial resolution is diffraction-limited typically to about a fraction of micrometer in the visible, and time resolution could be, depending on the hardware, as low as several picoseconds[24]. The difference between the ixFLIM and standard FLIM lies in the form of the excitation. Both in FLIM and ixFLIM, the excitation needs to be pulsed to resolve the decay in time. However, in ixFLIM the pulses must be broadband to spectrally cover the absorption of all fluorophores of interest, and their spectrum is modulated. Based on an approach well known from Fourier transform spectroscopy ixFLIM uses spectral interference of two identical pulses with variable time delay $\tau$ between them that leads to a cosine spectral modulation at each delay time[23,25] (Fig. 1b). For each delay $\tau$ of the excitation pulses, a frame of FLIM image is recorded. The total ixFLIM data are thus in form of a four-dimensional dataset $\text{ixFLIM}_{\text{raw}}(x, y, t, \tau)$, where (x,y) are the coordinates of the 2D image plane, $t$ is the TCSPC time delay and $\tau$ the excitation-pulse delay. The details of the signal processing as well as the implementation of the experiment in our laboratory are described in detail in the Methods section. Briefly, the dataset is Fourier-transformed along $\tau$, yielding the four-dimensional hypercube $\text{ixFLIM}_{\text{raw}}(x, y, t, \omega_\tau)$. After division by the known imprinted excitation spectrum along $\omega_\tau$, we obtain the ixFLIM signal $\text{ixFLIM}(x, y, t, \omega_\tau)$, resolved along the excitation frequency $\omega_\tau$ and emission time $t$. At each pixel (coordinates x,y) of the image, the correlation of the excitation spectrum and fluorescence decay is thus obtained (Fig. 1c). Importantly, the excitation wavelength is not scanned sequentially, but all wavelengths are present in each pulse, with varying amplitude. It is worth mentioning that the ixFLIM dataset contains the standard FLIM as its marginal, $\text{FLIM}(x, y, t) = \int \text{ixFLIM}(x, y, t, \omega_\tau) d\omega_\tau$. The marginal in along the other dimension is the excitation-spectrum resolved, interferometric excitation fluorescence image (ixFIM): $\text{ixFIM}(x, y, \omega_\tau) = \int \text{ixFLIM}(x, y, t, \omega_\tau) dt$.

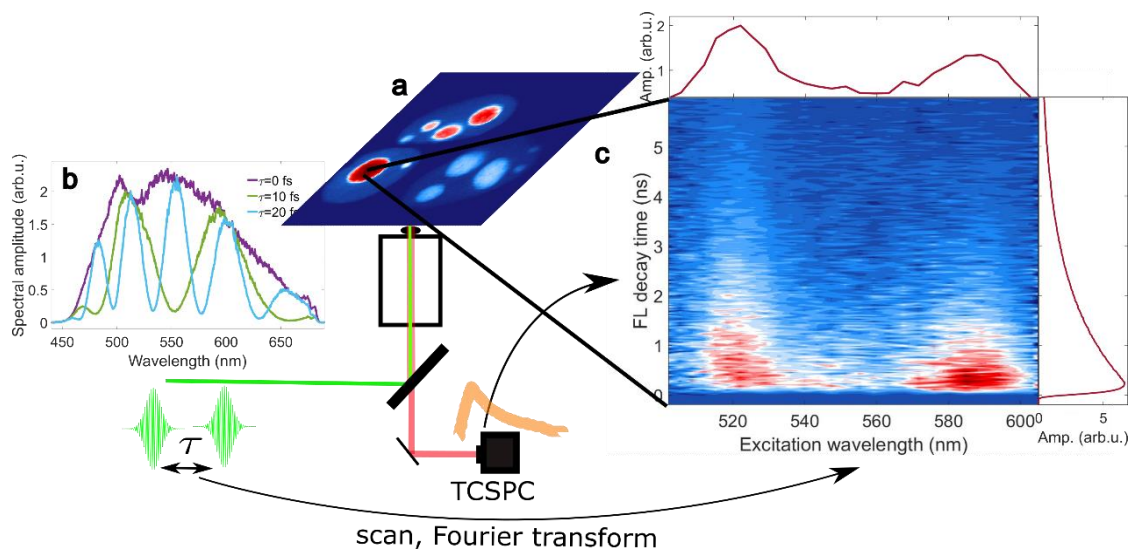

**Figure 1 | Principle of ixFLIM.** a) Standard time-domain FLIM with point scanning and TCSPC detection. b) Spectral modulation of broadband excitation pulses by interference of two pulses delayed by $\tau$. c) Fourier transform of the time resolved FLIM data leads to the transient map at that correlates the fluorescence excitation spectrum (horizontal) and emission decay (vertical) at each pixel of the image. The marginals of the transient map represent the total fluorescence decay (vertical) and excitation spectrum (horizontal).

*Information in ixFLIM: Oxonol VI binding to albumin*

To demonstrate how ixFLIM works, we measured fluorescence dye Oxonol VI (Oxo VI) in the presence and absence of bovine serum albumin (BSA). Oxo VI is a cationic potential-sensitive redistribution probe that penetrates biological membranes and, according to the membrane potential, partitions between cell/organelle exterior and interior where it binds to cellular constituents such as lipid or proteins [26,27]. When bound, its absorption spectrum shifts to the red (Fig. S3 in the SI) and fluorescence lifetime prolongs[28]. The magnitude of these changes is often used empirically to assess the relative fraction of bound vs free fluorophore.

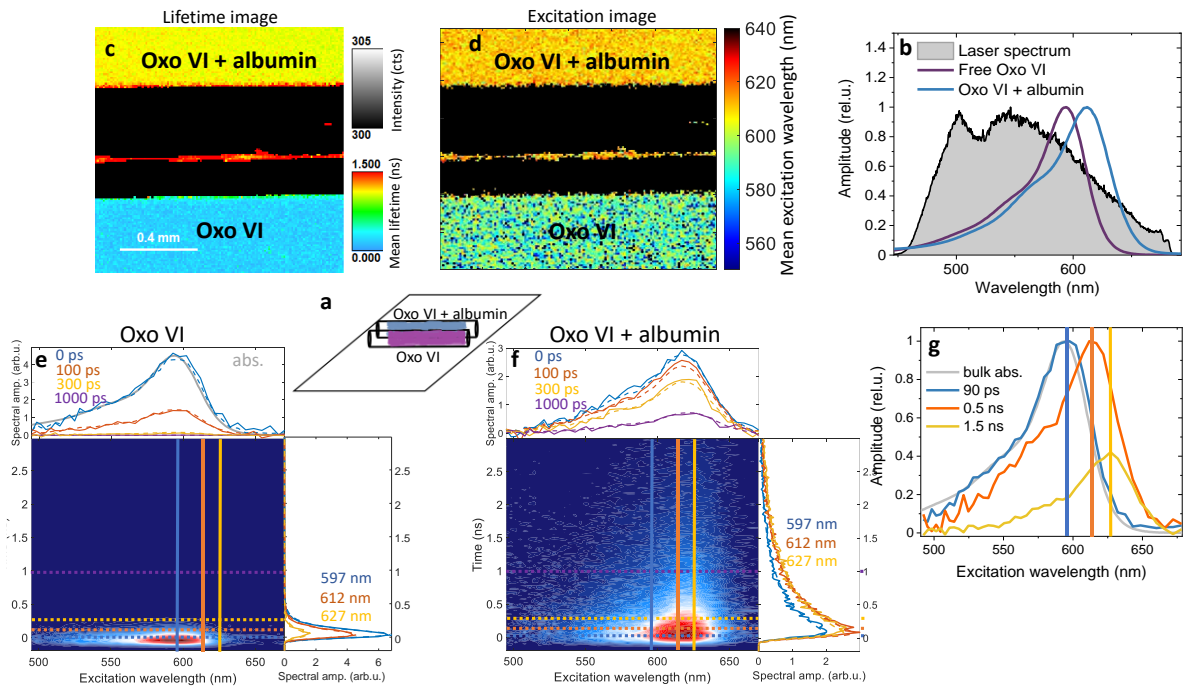

**Figure 2 | ixFLIM measurement of Oxonol VI.** a) sketch of the microcapillaries filled with free (purple) and bound (turquoise) Oxo VI. b) Bulk absorption spectra of free (purple) and bound (turquoise) oxonol, covered by the broadband excitation spectrum (black). c) FLIM image $FLIM(x, y, \langle t \rangle)$ false-colored by mean lifetime. d) fluorescence excitation spectrum image $ixFIM(x, y, \langle \omega_\tau \rangle)$ false-colored by mean excitation wavelength. e) Transient map $ixFLIM(t, \omega_\tau)$ of free Oxo VI (from the bottom capillary), with spectral and temporal cuts at the marked positions, overlaid by the global analysis fit (dashed lines). f) Same as in e) but for the upper capillary with bound Oxo VI. g) Decay-associated spectra of the time components from the global analysis of the free Oxo VI (blue, compared to its absorption spectrum in grey), and Oxo VI bound to albumin (red and yellow). The global analysis has been done on the maps in e) and f), and the position of the corresponding vertical lines marks the maxima of the transient spectral components.

We imaged two adjacent microcapillaries, one filled with free Oxonol VI in a PBS buffer and the other with Oxo VI bound to albumin (Fig. 2a), with absorption of both the bound and free Oxo VI covered by the broadband excitation laser spectrum (Fig. 2b). The difference between the two forms is visible in a standard FLIM, which is obtained from the ixFLIM dataset by integration over $\omega_\tau$. The mean (first moment) lifetime, $\text{FLIM}(x, y, \langle t \rangle)$, can be used for FLIM imaging without any fitting, distinguishing the bound and unbound oxonol clearly (Fig. 2c). The direct analogy in the spectral domain is to image by the mean (center of mass) excitation wavelength, $\text{ixFIM}(x, y, \langle \omega_\tau \rangle)$, obtained from the excitation-spectrum image ixFIM, as shown in Fig. 2d. Clearly, the regions of bound and unbound Oxo VI can be distinguished both by the mean fluorescence lifetime and by the mean excitation wavelength. However, there is much more information present in the ixFLIM dataset. For two regions of interest coinciding with the two micro-capillaries, we show the transient excitation maps obtained by data integration over x and y in Fig. 2e (free Oxo VI) and Fig. 2f (bound Oxo VI). Common techniques such as global analysis can be used to decompose the transient map into the individual species, obtaining their spectra and decay times (dashed fit lines in Fig. 2e,f, and the component spectra in Fig. 2g). In our case, the free Oxo VI is a single species with 90 ps lifetime and absorption peaking around 595 nm (blue in Fig. 2f), in agreement with literature[26,28,29]. In contrast, the Oxo VI bound to BSA is present in two species with lifetimes of 0.5 ns and 1.5 ns, with spectra progressively red-shifted to 620 nm and 632 nm, respectively (red and yellow in Fig. 2f). This correlates perfectly with the two known major binding sites of albumin[30], which apparently lead to a different magnitude of the spectral shift, correlating with lifetime prolongation. The presence of multiple spectra and kinetics is important when using Oxo VI as a membrane probe. While we used Oxo VI and albumin, the same principle (spectral and lifetime shifts, multiple binding sites) applies to most of other proteins and fluorophores. Beyond merely illustrating the principle of ixFLIM, this example highlights its utility in disentangling different species. Importantly, the same transient map is contained in each pixel of the image. As we will demonstrate in further examples, one can use the component separation to image the individual species.

*FRET-ixFLIM molecular ruler: Cy3-Cy5 at DNA*

The Oxo VI example served to illustrate the type of information contained in the ixFLIM dataset. We proceed with application to Förster Resonance Energy Transfer (FRET)[31], which is commonly used to measure nanometer-scale distances in biostructures, as well as to evaluate interaction between labelled biopolymers[1,12]. The excitation energy transfer rate $k_T$ between donor and acceptor fluorophores distance $R$ apart is

$$k_T = k_D \left(\frac{R_0}{R}\right)^6, \quad (1)$$

where $k_D$ is the fluorescence decay rate of the donor and $R_0$ is a so-called Förster critical radius, which can be expressed as[3,32]

$$R_0^6 = 8.79 \cdot 10^{-5} \left[\hat{\mu}_a \hat{\mu}_d - 3(\hat{\mu}_a \cdot \hat{R})(\hat{\mu}_d \cdot \hat{R})\right]^2 n^{-4} \int d\lambda \, \alpha_a(\lambda) \phi_d(\lambda) \lambda^4. \quad (2)$$

Here, $\alpha_a(\lambda)$ and $\phi_d(\lambda)$ are the absorption spectrum of the acceptor, and normalized emission spectrum of the donor, respectively. $\hat{\mu}_{a,d}$ are the acceptor and donor transition dipole moments normalized to unity, $\hat{R}$ is the normalized donor-acceptor distance, and *n* is the refractive index. While the $\frac{1}{R^6}$ dependence allows the distance-rule function, the orientation factor in the square brackets provides sensitivity to the dipole geometry. The transfer efficiency *E* is given by the competition of the transfer and donor excited state decay:

$$E = \frac{k_T}{k_T + k_D} = \frac{1}{1+\left(\frac{R_0}{R}\right)^6}. \quad (3)$$

To demonstrate the capability of quantitative FRET-ixFLIM imaging, we measured specially designed internally cyanine-labelled DNA oligonucleotides[33] (Fig. 3a). The donor Cy3 and acceptor Cy5 dyes were covalently integrated to the sugar-phosphate backbone of complementary DNA single strands at precisely determined positions (see Methods). Upon hybridization, these formed a double-stranded DNA with donor acceptor pairs at four precisely defined positions, at distances of 4 base pairs (bp04), 12 bp (bp12), 13 bp (bp13) and 20 bp (bp20). Cyanine dye labelling has been previously used to study the local DNA conformation, leveraging the FRET sensitivity[34]. It thus remains important to know the extent of the intrinsic geometrical disorder in these structures.

For imaging purposes, the DNA constructs were attached by a biotin-streptavidin link to magnetic beads (2.8 µm in diameter) with the size comparable to cellular structures, such as nucleoli[11,35]. Such beads can be easily imaged by the fluorescence microscope, (Fig. 3b, Cy3–Cy5 distance 20 base pairs). A standard FLIM image shows a uniform narrow distribution of lifetimes around 1.4 ns (inset in Fig. 3b), consistent with that of the Cy5 acceptor[36], with small deviations easily explained by the known sensitivity of the fluorescence to the particular position along the DNA[37,38]. Our broadband excitation laser spectrum comfortably covers absorption of both the Cy3 and Cy5 (Fig. 3c). The ixFLIM transient map obtained from all the beads in Fig. 3b is shown in Fig. 3d. Two clear absorption bands are visible, corresponding to the donor and acceptor spectra. Since Cy3 and Cy5 are well spectrally separated and we use a 700 nm longpass filter in the detection path, we detect almost exclusively emission from the Cy5 acceptor. This can be verified by checking that at longer delay times, signal decays exclusively with the acceptor lifetime. The presence of the donor Cy3 peak in the ixFLIM spectrum is in this case immediately a qualitative proof that the FRET takes place.

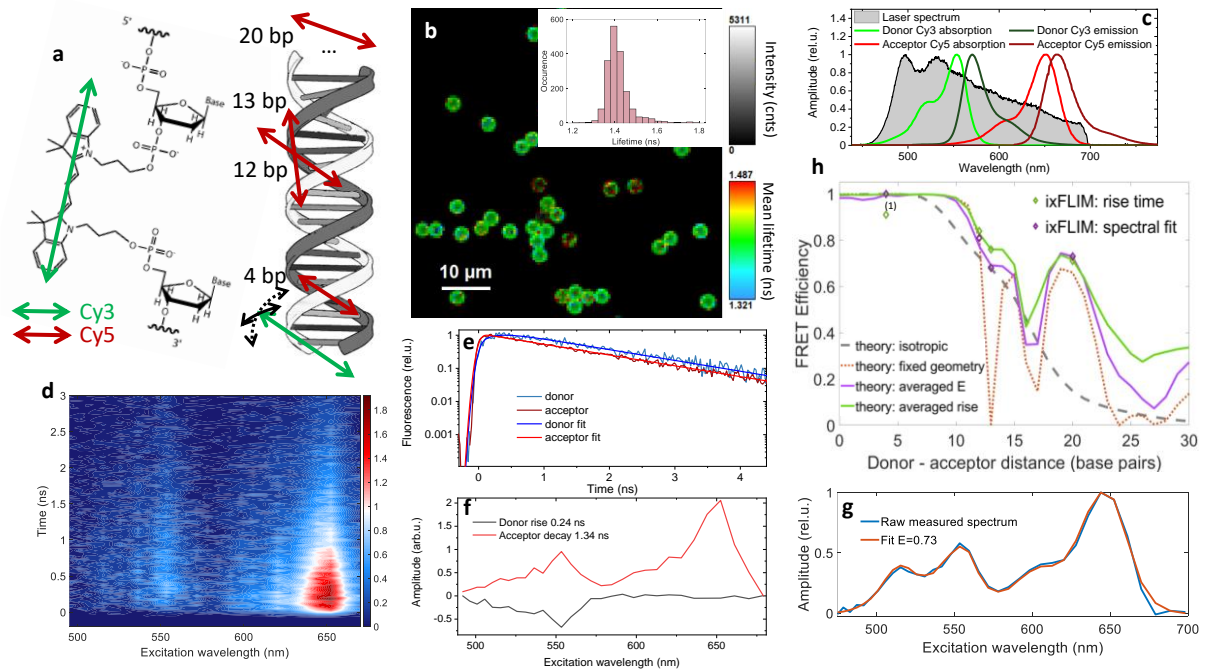

**Figure 3 | ixFLIM-FRET demonstration on internally labelled Cy3-Cy5 DNA constructs.** a) Geometry of the dye attachment to the DNA (picture adapted from Ref. 33. The double-sided red and green arrows mark the Cy3 and Cy5 transition dipoles, the numbers indicate their mutual distance in base pairs (bp). The black arrows indicate the considered disordered motion. b) FLIM $FLIM(x, y, \langle t \rangle)$ image of the beads to which the DNA constructs are attached with 20 bp Cy3-Cy5 distance. Lifetime is narrowly distributed around 1.4 ns (inset). c) Cy3 (green) and Cy5 (red) absorption (light shade) and emission (dark shade) spectra, overlaid by the broadband excitation pulse spectrum (black), truncated by a 700 nm short-pass filter to facilitate detection of longer-wavelength Cy5 emission. d) ixFLIM transient map obtained from the image in b). e) Transients in the donor and acceptor spectral region (as identified spectrally according to c), overlaid with fits from global analysis. f) Spectral components from the global analysis, 240 ps rise of the donor signal due to FRET (black), 1.34 ns decay of the emitting acceptor (red). g) Spectral fit (orange) to extract *E* according to Eq. (5), decomposing the raw excitation spectrum (blue, not divided by the laser spectrum) into the weighed sum of the donor and acceptor absorption spectra (multiplied by the excitation laser spectrum). h) Dependence of the FRET efficiency on the donor-acceptor position along the DNA helix, comparison of theory and experiment. (1) indicates that the fastest FRET could not be reliably fit from the signal rise, see also the SI.

There are two ways to quantify the FRET efficiency. First, from the signal kinetics. Since the excitation of the donor can lead to acceptor emission (with rate $k_A$) only after the energy is transferred, the donor signal appears with a delay with respect to the acceptor fluorescence. As derived in the Methods, the FRET-ixFLIM signal detected by acceptor fluorescence can be expressed as

$$\text{ixFLIM}_A(\omega, t) = k_A \phi_A \left\{ c_A \epsilon_A(\omega) e^{-k_A t} + c_D \epsilon_D(\omega) \frac{k_T}{k_T + k_D - k_A} \left( e^{-k_A t} - e^{-(k_T + k_D)t} \right) \right\}. \quad (4)$$

Here, $\phi_A$ is the acceptor fluorescence quantum yield, $c_{A,D}$ are the acceptor and donor concentrations (in our case of Cy3-Cy5 pairs we have $c_A = c_D$ by design) and $\epsilon_{A,D}(\omega)$ are the excitation spectra of the donor and acceptor, respectively. There are therefore just two time components in the kinetics: one that contains both donor and acceptor spectra and decays with the lifetime of the acceptor, and a second one that rises with the sum of the donor decay rate and the FRET transfer rate[39]. If the donor decay rate is known, e.g., from an independent measurement, the FRET rate $k_T$ can thus be easily determined. This is demonstrated in Fig. 3e, where the fits of the transients at the peak positions of the donor and acceptor clearly show the delayed rise of the donor peak due to the FRET. The fitting was done on the whole transient map using global analysis, in which the two components have been identified (Fig. 3f). The first one (black line) has negative spectral amplitude signifying signal rise and spectrum of the donor (compare to Fig. 3c). This is thus the component connected to the FRET, with the rise time of $\tau_\text{rise} = (k_D + k_T)^{-1} = 235$ ps. Considering the known donor lifetime of 0.82 ns (see the measurement in the SI), we obtain transfer time of 0.33 ns and, thus, FRET efficiency $E = 0.71$.

A second way to quantify the FRET efficiency is from the time-integrated excitation spectrum $\text{ixFIM}(x, y, \omega_\tau)$ (see Methods section):

$$\text{ixFIM}_A(\omega) = \int \text{ixFLIM}_A(\omega, t) dt = \phi_A c_A \{ \epsilon_A(\omega) + \epsilon_D(\omega) E \}. \quad (5)$$

The contribution of the donor relative to the absorption spectrum determines the transfer efficiency $E$. For this, the absorption spectra of the donor and acceptor have to be known, as is the case for most commonly-used FRET probes[40] (Fig. 3c and the SI). The example of the fit of the excitation spectra obtained from the time-integrated transient map is given in Fig. 3g, getting the efficiency of E=0.73. Interestingly, as we demonstrate in the SI, the difference between the *E* recovered from the spectral and temporal fit does not stem from uncertainty in the fitting but is rather a consequence of a different quantity (signal rise and donor–acceptor excitation spectrum) being averaged over the distributions of the fluorophore geometry in the fitted experimental data. Note that the two approaches are to a degree complementary. Fast, efficient transfer is difficult to discern in the time domain, but the donor spectrum will be present strongly so that the spectral fit is reliable. On the other hand, slow inefficient transport can be conveniently tracked in time domain, where spectral integration over the donor peak will increase the signal to noise ratio. In between, there is a wide region of transfer efficiencies where both approaches work well. There, the combination of the two approaches can be used to eliminate an additional unknown parameter, such as unknown change in spectrum or lifetime. In this sense, the FRET-ixFLIM has a strong intrinsic reference that makes it more robust than standard FLIM-FRET.

In Fig. 3b-g, the distance between the donor and acceptor is fixed to 20 base pairs. With the two ways to determine the FRET efficiency from the ixFLIM data established, we varied the donor-acceptor distance along the DNA helix in order to study the FRET as a function of distance and geometry, similar to Ref. [41]. The donor (Cy3) position was fixed, and the acceptor Cy5 increased from 4 bp to 12 bp, 13 bp and 20 bp. (Fig. 3a and Methods). Due to their rigid integration to the helix backbone, not only the fluorophore distance but also their mutual orientation changes with the position along the helix. However, due to inevitable structural disorder, there must be a distribution of the dye orientations, which is a-priori unknown[41]. Knowing the Förster radius of the Cy3-Cy5 pair $R_0 = 5.3$ nm[40], the theoretical dependence of the FRET efficiency $E$ on the position can be calculated, as acquired from the spectra and signal rise. In Fig. 3h we plot these dependencies for random (isotropic) fluorophore orientation along the chain, as would be the case for external labelling (grey), for their fixed orientation with dipole moments locally pointing along the DNA strands (orange), and for the realistic situation with relaxed DNA rigidity allowing some structural dipole disorder (green and purple). In the calculation, the DNA geometry was fixed according to that of B-type DNA and the extent of the dye orientational disorder and the precise distance of the fluorophore transition dipole from the DNA axis were taken as fit parameters and optimized by a global least-square algorithm to fit experimental data (see SI). The results of the calculation, as well as the experimental ones, are summarized in Table 1, and compared visually in Fig. 3g.

*Table 1* FRET efficiency as a function of Cy3-Cy5 position along the DNA: summary of results. Same as in Fig. 3h, (1) indicates that the fastest FRET could not be reliably fit from the signal rise.

| Cy3-Cy5 distance | Signal rise fit $k_T^{-1}, E_\text{rise}$ | Theory: $\langle \frac{k_T}{k_T+k_D-k_A} e^{-(k_T+k_D)t} \rangle$ | Spectral fit | Theory: $\langle E \rangle = \langle \frac{k_T}{k_T+k_D} \rangle$ |
|---|---|---|---|---|
| 4 bp (1.36 nm) | 80 ps$^{(1)}$, 0.91 | 1.0 | 1.0 | 1.0 |
| 12 bp (4.08 nm) | 170 ps, 0.83 | 0.83 | 0.81 | 0.77 |
| 13 bp (4.42 nm) | 260 ps, 0.76 | 0.76 | 0.68 | 0.69 |
| 20 bp (6.8 nm) | 330 ps, 0.71 | 0.72 | 0.73 | 0.73 |

Clearly, the agreement between the experiment and theory is very good, verifying the quantitative performance of FRET-ixFLIM. The non-monotonous dependence of the transfer efficiency $E$ on the donor-acceptor distance indicates the sensitivity of FRET-ixFLIM to the dipole geometry[41]. The significant difference of the transfer efficiency from the fixed-geometry values (seen, e.g., as a deep dip at position 12 with on-average near-perpendicular dye transition dipoles) furthermore indicates the sensitivity to the structural disorder. The extracted values of no disorder in the helical rise angle, 27 deg (FWHM) disorder in the azimuthal angle and 10.6 Å distance of the dye transition dipole from the helical axis serve to characterize the intrinsically labelled DNA-cyanine dye constructs. It might be surprising that, despite the expectedly rigid intrinsic labelling and rigid dsDNA fragment significantly shorter than the expected persistent length (~150 bp)[42] the dye orientation remains somewhat flexible. This may be useful for studies of the local DNA geometry[34] and flexibility[43].

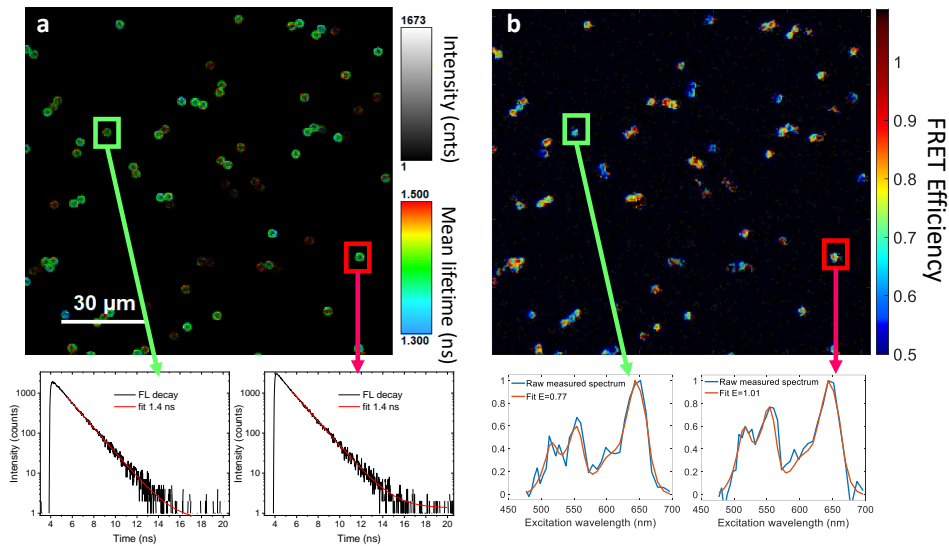

**Figure 4 | *FRET*-ixFLIM imaging.** a) FLIM image of a mixture of beads with Cy3-Cy5 donor-acceptor pairs separated by 4 bp and 20 bp b) The same image, but false-colored by FRET efficiency extracted in each pixel from the ratio of the donor and acceptor excitation spectra. Two beads are highlighted in both images, with nearly identical lifetime, and very different FRET efficiency of E=0.77 (green, 20 bp) and E=1 (red, 4 bp).

For the accurate determination of the FRET efficiency, we used data averaged over about 10-30 beads (Fig. 3b for 20 bp and SI for 4, 12 and 13 bp). However, the same information, albeit noisier, is present in each pixel of the image. One can thus, for example, use the spectral method to determine *E* and use it to false-color the image by FRET efficiency. To demonstrate this, we prepared a mixture of bp04 and bp20 beads. As shown in Fig. 4, the beads have a similar lifetime (of the acceptor), but the FRET efficiency clearly distinguishes the two fractions of the beads, with efficient and less efficient FRET. Note, that the most efficient FRET (nearly 100%) would be impossible to observe in donor emission by standard FLIM only, since efficiency over 95% implies transfer time of less than 43 ps, which results in very dim quenched donor with the emission decay within 40 ps. This time is on the edge of time resolution for the acceptor-FLIM rise and decreases the donor emission to less than 5% in donor-FLIM. In contrast, FRET-ixFLIM can easily observe constructs with both fast and slow transfer.

*Protein interaction: Nucleophosmin in nuclei and nucleoli of HEK-293T cells*

In many biological and biophysical applications, FRET is used for detection of protein interactions and formation of macromolecular assemblies[10,12]. To demonstrate the applicability and advantage of ixFLIM in such situations, we imaged HEK-293T cells cotransfected with the abundant nucleolar protein nucleophosmin (NPM) tagged with donor mVenus or the acceptor mRFP1 (see Methods for sample preparation). With this dual labelling, one can study the oligomerization state of NPM in cell nuclei and nucleoli[11]. It is assumed that NPM forms higher oligomers, mainly pentamers or decamers in nuclei[44]. If that is the case, there should be a significant fraction of the mixed oligomers with donor-acceptor pairs and thus FRET present[11,45]. As before, we measured ixFLIM with excitation of both the donor and acceptor (Fig. 5a), and detection of the emission using a 655 nm longpass filter. Compared to the Cy3-Cy5 pairs, the situation is somewhat more complicated for the mVenus-mRFP1 pair. The donor mVenus has longer fluorescence lifetime (2.9 ns)[46] and higher quantum yield (0.64)[40] than the lifetime (1.8 ns)[47] and quantum yield (0.25)[40] of the acceptor. Their smaller spectral separation therefore results in both donor and acceptor emission present in the collected data. Furthermore, while in the DNA constructs we had always a donor-acceptor pair present, here we have donor and acceptor in different stoichiometries randomly present in the individual NPM oligomers, including some fractions of non-interacting donors and non-interacting acceptors only. Finally, NPM oligomers may contain non-labeled endogenous NPM. This real-world application is thus a good test for ixFLIM. In Fig. 5a and 5b the fluorescence image is shown of a HEK-293T cell nucleus with two nucleoli, colored by the donor average lifetime (Fig. 5b) and the excitation wavelength (Fig. 5c). Since the NPM localizes mainly in the nucleoli, these are much brighter than the surrounding nucleoplasm. The mean fluorescence lifetime is around 2.2 ns and the central absorption wavelength around 540 nm. The overall, spatially integrated and spectrally-unresolved fluorescence decay (Fig. S4 in the SI) can be described by three time components: 2.94 ns, 1.81 ns and 703 ps. The 2.94 ns one indicates the free mVenus donor[46], and 1.81 ns corresponds well to the expected lifetime of the mRFP1 acceptor[47]. But what is the additional component? How large is the free-donor fraction, and is there donor-acceptor interaction? These questions can be answered by the excitation frequency–decay time correlation in the transient map of ixFLIM shown in Fig. 5d.

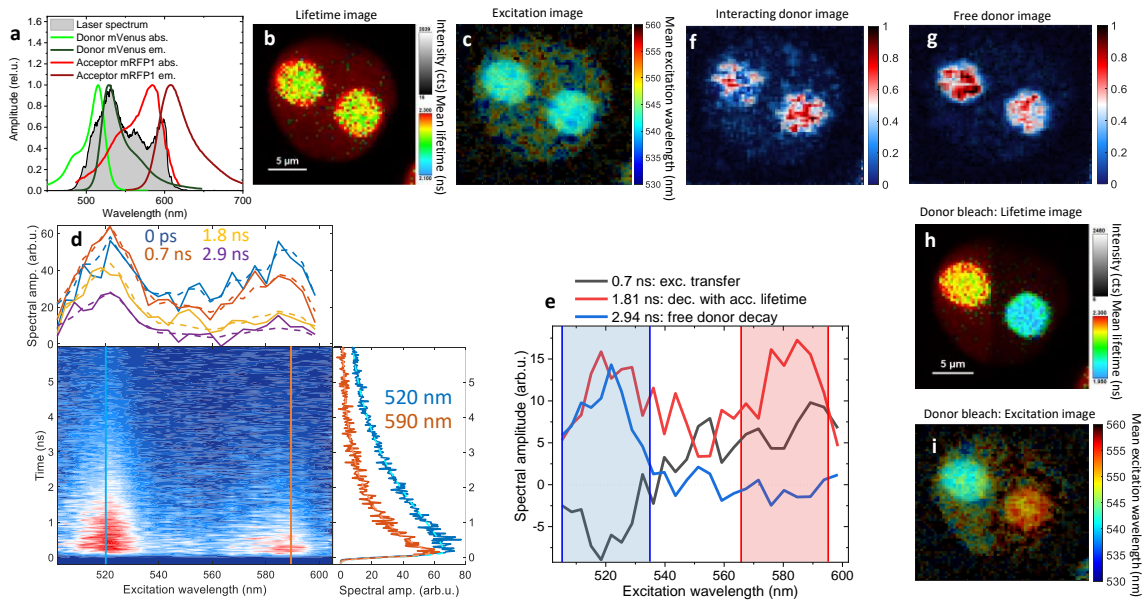

**Figure 5 | Interaction of NPM in nuclei of HEK-*293T* cells by ixFLIM.** a) mVenus (green) and mRFP1 (red) absorption (light shade) and emission (dark shade) spectra, overlaid by the broadband excitation pulse spectrum (black). Emission was detected through a 655 nm long-pass filter. b) Fluorescence image of a cell nucleus with two nucleoli, colored by the mean lifetime, $FLIM(x,y,\langle t \rangle)$. c) The same nucleus coloured by the mean excitation wavelength, $ixFIM(x,y,\langle \omega_\tau \rangle)$ d) ixFLIM transient map $ixFLIM(t,\omega_\tau)$ of the nucleus imaged in b) and c), with spectral cuts at the indicated times, and temporal cuts at the labelled positions. e) Spectra of the time components from the global analysis of the map in d), components associated with rates $k_T + k_D$ (black), $k_A$ (red) and $k_D$ (blue). Shaded rectangles mark the donor mVenus (blue) and acceptor mRFP1 (red) regions used for the analysis as described in the text. f) Image of the nucleus coloured by the amplitude of the emission from the interacting donor (normalized to maximum). g) Same as in f) but colored by the emission from the free donor. h) FLIM image, $FLIM(x,y,\langle t \rangle)$ of the same nucleus as in a) but with the right half of the nucleus bleached to 1/3 of the initial emission intensity by 488 nm light (mainly donor bleach). The donor bleaching shortens the detected mean emission lifetime which gets dominated by the shorter-lifetime acceptor emission. i) Excitation wavelength image $ixFIM(x,y,\langle \omega_\tau \rangle)$ of the same nucleus with bleached right half. The donor bleaching increases the mean excitation wavelength which gets dominated by the redder absorption of the acceptor.

As derived in the Methods section [Eq. (12)], the overall ixFLIM transient spectrum can be described as

$$ixFLIM(\omega_\tau, t) = k_A \phi_A \left\{ c_A \epsilon_A(\omega_\tau) e^{-k_A t} + c_D^b \epsilon_D(\omega_\tau) \frac{k_T}{k_T + k_D - k_A} \left( e^{-k_A t} - e^{-(k_T + k_D)t} \right) + \right.$$

$$\left. \frac{k_D \phi_D}{k_A \phi_A} \epsilon_D(\omega_\tau) \left( c_D^f e^{-k_D t} + c_D^b e^{-(k_T + k_D)t} \right) \right\}. \quad (6)$$

Same as before, here, $k_A, k_D, k_T$ are the acceptor and donor decay rates and the transfer rate as before, $c_A, c_D^b, c_A^f$ are the concentrations of acceptor, bound and free donor, $\phi_{A,D}$ are the respective quantum yields that include also detection efficiency, and $\epsilon_A(\omega_\tau), \epsilon_D(\omega_\tau)$ are the donor and acceptor excitation spectra. One therefore should expect three time components in the data. Performing global analysis, we find that the whole transient map can indeed be described by the three time constants given above. The comparison of the fit to the data is shown in the spectral and temporal cuts to the transient decay map in Fig. 5d. The spectra associated to the decay components (Fig. 5e) can be decomposed into the donor and acceptor contributions (the respective regions are marked in blue for donor and red for acceptor in Fig. 5e). The longest time component of $k_D^{-1}$ =2.94 ns, features, as expected, the excitation spectrum of the pure donor (compare to Fig. 5a). The $k_A^{-1}$ =1.81 ns component features both donor and acceptor spectra, i.e., proves that there is FRET and thus NPM oligomerization takes place. The 703 ps component features a small contribution of the acceptor decay, and, crucially, a rise of the donor signal. This time constant thus corresponds to $(k_T + k_D)^{-1}$, same as for the Cy3-Cy5 DNA constructs. However, here we have the free donor present within the same measurement, with $k_D = \frac{1}{2.94 \text{ ns}}$. We therefore immediately obtain the transfer time of 0.92 ns, and thus transfer efficiency of $E = 0.24$ (Eq. (3) above). Assuming random dye orientation, this implies distance of ~66 Å between the dyes (critical radius of mVenus–mRFP1 pair is 55Å)[40]. This is consistent with the oligomeric geometry, considering the NPM size of about 50Å[44].

Furthermore, from the separation into components the fraction of the free and bound donor can be extracted as follows. According to the expression for the spatially-integrated $ixFLIM(\omega_\tau, t)$ above, the free donor component $(k_D)$ has a relative amplitude of $A_D^{(k_D)} = \frac{k_D \phi_D}{k_A \phi_A} c_D^f$, while the bound donor part decaying with the acceptor lifetime $(k_A)$ has amplitude of $A_D^{(k_A)} = \frac{k_T}{k_T + k_D - k_A} c_D^b$, and the bound donor rising component $(k_D + k_T)$ has an amplitude of $A_D^{(k_D + k_T)} = -c_D^b \left( \frac{k_T}{k_T + k_D - k_A} - \frac{k_D \phi_D}{k_A \phi_A} \right)$. These amplitudes can be directly extracted from their associated spectra (Fig. 5e), identifying the acceptor (565 nm–595 nm) and donor (505 nm–535 nm) absorption regions, with a correction for the 23% contribution of the acceptor absorption in the donor region[40]. First, we use $\frac{A_D^{(k_A)}}{A_D^{(k_D + k_T)}} = \frac{A_{505-535}^{1.81 \text{ ns}} - 0.23 A_{565-595}^{1.81 \text{ ns}}}{A_{505-535}^{0.7 \text{ ns}}}$ to extract the unknown $\frac{k_D \phi_D}{k_A \phi_A} = 0.59$. Then, we use $\frac{A_D^{(k_D)}}{A_D^{(k_A)}} = \frac{A_{505-535}^{2.94 \text{ ns}}}{A_{505-535}^{1.81 \text{ ns}} - 0.23 A_{565-595}^{1.81 \text{ ns}}}$ to obtain $\frac{c_D^f}{c_D^b} = 1.68$. We therefore obtain the fraction of interacting donor $f_D^{int} = \frac{c_D^b}{c_D^f + c_D^b} = 0.37$. Because the NPM concentration is much higher in the nucleoli than in the surrounding nucleoplasm, the fraction of the bound donor is expected to be higher in the nucleoli as well. We thus perform the same analysis separately for the two compartments (see the SI), obtaining $f_D^{int}|_{nucleoplasm} = 0.21$, while $f_D^{int}|_{nucleoli} = 0.44$. If one would measure the NPM interaction by donor mVenus, the average donor lifetime (calculated as the first time moment of the decay) would thus shorten from $k_D^{-1}$ =2.94 ns to $(k_D + k_T)^{-1}$ =2.58 ns. This is indeed what was observed in the previous experiments[11].

Since the component decaying with the acceptor lifetime and the excitation spectrum of the donor directly reports on the interaction, we can use its amplitude for imaging of the oligomerizing NPM. We thus isolate the 1.81 ns component for each pixel and each frame in the $\tau$ –shift stack, and then integrate over the donor absorption in the spectral domain, correcting for the acceptor contribution. The result, shown in Fig. 5f, is the direct image of the interacting mVenus donor, and thus oligomerized NPM. It can be compared to the non-transferring donor (Fig. 5g).

Finally, to support further our assignment of the spectral components and demonstrate the interaction-based imaging, we bleach the right half of the nucleus with intense 488 nm light, which predominantly (77%) bleaches the donor[40] (see spectra in Fig. 5a). The resulting image in average acceptor lifetime (Fig. 5h), average spectral wavelength (Fig. 5i) fits very well our picture. Upon donor photobleaching, the average measured lifetime shortens since the component with the longer lifetime of the donor is suppressed in the detected signal, and the spectrum red-shifts since the blue donor absorption is missing.

To summarize, FRET-ixFLIM allows direct, information-rich quantitative imaging of *in vivo* intracellular interactions within a single experiment without need of tedious control bleaching experiments which could bias result e.g. by progressive sample evolution or degradation or by unwanted spectral photoconversion, which often may happen for fluorescent protein tags[2,13].

**Discussion**

We have introduced a new variant of FLIM with interferometric excitation, that we call ixFLIM. ixFLIM correlates the excitation spectrum and emission lifetime in each pixel (or voxel in case of 3D scanning) of the fluorescence image. As we have shown, the additional dimension provides FLIM with capability to assign and resolve various species in the fluorescence image. One of the most promising applications of ixFLIM is to the FRET measurement. We have established a methodology using FRET-ixFLIM to quantitatively measure distances and geometry of the interacting molecular pairs, and to prove, isolate and image protein interaction within a single measurement. With ixFLIM, the FRET species can be reliably identified, and there is no need for acceptor bleaching to prove the FLIM. Detection of the acceptor instead of the donor allows a different use of FRET, where the acceptor emission can be used to detect the donor excitation spectrum, even for donors with small quantum yield or emission in unfavorable spectral region.

As briefly mentioned in the introduction, related techniques exist in the literature. The closest one is that recently applied by Thyrhaug et al. to single-molecules[23]. Using the same type of interferometric two-pulse excitation as in this work, they resolved the molecular excitation spectra, which they correlated with emission spectra and, alternatively, with the fluorescence decay by TCSPC. The decay was, however, the same across the whole spectral range and thus no information beyond the two one-dimensional datasets (excitation spectrum and fluorescence decay) was present.

When extending a standard FLIM to ixFLIM, only the excitation has to be modified. ixFLIM can thus be implemented in practically any existing FLIM microscope or imager (time or frequency domain, scanning or wide field), extending its capabilities. Regarding sample stability and acquisition time, ixFLIM is comparable to standard FLIM. In practice, the excitation spectrum is modulated by stepping the delay time $\tau$ during the acquisition of the FLIM frames, which are normally acquired sequentially for averaging. In the measurements presented in this manuscript, the typical scanning time per frame was about 20 $\mu$s per pixel, i.e., about 1.3 s for 256x256 pixel image, and 301 frames were acquired, which corresponds to a measurement time of about 7 min. The excitation light intensity is the same as used in the standard FLIM, the limitations being pile-up in the photon counting and sample bleaching. These measurement parameters are fully compatible with *in vivo* cell imaging.

While this work showcases several biophysical applications of ixFLIM, other possible application areas span all those where standard FLIM can be applied. Examples include biomedical research but also optoelectronics and photovoltaics, forensics, sensing or agricultural applications. We are convinced that ixFLIM will prove to be useful in a wide range of applications in life sciences and beyond.

**References**

1. Ouyang, Y., Liu, Y., Wang, Z. M., Liu, Z. & Wu, M. FLIM as a Promising Tool for Cancer Diagnosis and Treatment Monitoring. *Nano-Micro Lett.* **13**, 1–27 (2021).

2. Herman, P., Holoubek, A. & Brodska, B. Lifetime-based photoconversion of EGFP as a tool for FLIM. *Biochim. Biophys. Acta BBA - Gen. Subj.* **1863**, 266–277 (2019).

3. Lakowicz, J. R. *Principles of Fluorescence Spectroscopy*. (Springer, 2006).

4. Berezin, M. Y. & Achilefu, S. Fluorescence Lifetime Measurements and Biological Imaging. *Chem. Rev.* **110**, 2641–2684 (2010).

5. van Munster, E. B. & Gadella, T. W. J. Fluorescence Lifetime Imaging Microscopy (FLIM). in *Microscopy Techniques: -/-* (ed. Rietdorf, J.) 143–175 (Springer, 2005). doi:10.1007/b102213.

6. Chorvat Jr., D. & Chorvatova, A. Multi-wavelength fluorescence lifetime spectroscopy: a new approach to the study of endogenous fluorescence in living cells and tissues. *Laser Phys. Lett.* **6**, 175–193 (2009).

7. Alexiev, U., Volz, P., Boreham, A. & Brodwolf, R. Time-resolved fluorescence microscopy (FLIM) as an analytical tool in skin nanomedicine. *Eur. J. Pharm. Biopharm.* **116**, 111–124 (2017).

8. Suhling, K. *et al.* Imaging the Environment of Green Fluorescent Protein. *Biophys. J.* **83**, 3589–3595 (2002).

9. Pp, P., Kw, E. & Pj, K. Multiphoton microscopy and fluorescence lifetime imaging microscopy (FLIM) to monitor metastasis and the tumor microenvironment. *Clin. Exp. Metastasis* **26**, (2009).

10. Sun, Y., Hays, N. M., Periasamy, A., Davidson, M. W. & Day, R. N. Chapter nineteen - Monitoring Protein Interactions in Living Cells with Fluorescence Lifetime Imaging Microscopy. in *Methods in Enzymology* (ed. conn, P. M.) vol. 504 371–391 (Academic Press, 2012).

11. Strachotová, D., Holoubek, A., Wolfová, K., Brodská, B. & Heřman, P. Cytoplasmic localization of Mdm2 in cells expressing mutated NPM is mediated by p53. *FEBS J.* **290**, 4281–4299 (2023).

12. Jares-Erijman, E. A. & Jovin, T. M. FRET imaging. *Nat. Biotechnol.* **21**, 1387–1395 (2003).

13. Kremers, G.-J., Hazelwood, K. L., Murphy, C. S., Davidson, M. W. & Piston, D. W. Photoconversion in orange and red fluorescent proteins. *Nat. Methods* **6**, 355–358 (2009).

14. Popleteeva, M. *et al.* Fast and simple spectral FLIM for biochemical and medical imaging. *Opt. Express* **23**, 23511–23525 (2015).

15. Perri, A. *et al.* Time- and frequency-resolved fluorescence with a single TCSPC detector via a Fourier-transform approach. *Opt. Express* **26**, 2270–2279 (2018).

16. Becker, W., Braun, L., Jubghans, C., Bergmann, A. & Becker& Hickl GmbH. FLIM with Excitation-Wavelength Multiplexing. (2020).

17. Zhao, M., Huang, R. & Peng, L. Quantitative multi-color FRET measurements by Fourier lifetime excitation-emission matrix spectroscopy. *Opt. Express* **20**, 26806 (2012).

18. Zhao, M., Li, Y. & Peng, L. Parallel excitation-emission multiplexed fluorescence lifetime confocal microscopy for live cell imaging. *Opt. Express* **22**, 10221–10232 (2014).

19. Hybl, J. D., Albrecht Ferro, A. & Jonas, D. M. Two-dimensional Fourier transform electronic spectroscopy. *J. Chem. Phys.* **115**, 6606 (2001).

20. Tekavec, P. F., Lott, G. A. & Marcus, A. H. Fluorescence-detected two-dimensional electronic coherence spectroscopy by acousto-optic phase modulation. *J. Chem. Phys.* **127**, 214307 (2007).

21. Piatkowski, L., Gellings, E. & van Hulst, N. F. Broadband single-molecule excitation spectroscopy. *Nat. Commun.* **7**, 10411 (2016).

22. Fersch, D. *et al.* Single-Molecule Ultrafast Fluorescence-Detected Pump–Probe Microscopy. *J. Phys. Chem. Lett.* **14**, 4923–4932 (2023).

23. Thyrhaug, E. *et al.* Single-molecule excitation–emission spectroscopy. *Proc. Natl. Acad. Sci.* **116**, 4064–4069 (2019).

24. Becker, H. *The bh TCSPC Handbook. 10th edition, 2023*. (Becker and Hickel, GmBH).

25. Perri, A. *et al.* Excitation-emission Fourier-transform spectroscopy based on a birefringent interferometer. *Opt. Express* **25**, A483–A490 (2017).

26. Smith, J. C. & Chance, B. Kinetics of the potential-sensitive extrinsic probe oxonol VI in beef heart submitochondrial particles. *J. Membr. Biol.* **46**, 255–282 (1979).

27. Holoubek, A., Večeř, J., Opekarová, M. & Sigler, K. Ratiometric fluorescence measurements of membrane potential generated by yeast plasma membrane H+-ATPase reconstituted into vesicles. *Biochim. Biophys. Acta BBA - Biomembr.* **1609**, 71–79 (2003).

28. Bashford, C. L., Chance, B., Smith, J. C. & Yoshida, T. The Behavior of Oxonol Dyes in Phospholipid Dispersions. *Biophys. J.* **25**, 63–85 (1979).

29. Smith, J. C., Hallidy, L. & Topp, M. R. The behavior of the fluorescence lifetime and polarization of oxonol potential-sensitive extrinsic probes in solution and in beef heart submitochondrial particles. *J. Membr. Biol.* **60**, 173–185 (1981).

30. Nishi, K., Yamasaki, K. & Otagiri, M. Serum Albumin, Lipid and Drug Binding. in *Vertebrate and Invertebrate Respiratory Proteins, Lipoproteins and other Body Fluid Proteins* (eds. Hoeger, U. & Harris, J. R.) 383–397 (Springer International Publishing, 2020). doi:10.1007/978-3-030-41769-7_15.

31. Förster, Th. Zwischenmolekulare Energiewanderung und Fluoreszenz. *Ann. Phys.* **437**, 55–75 (1948).

32. Blankenship, R. E. *Molecular Mechanisms of Photosynthesis*. (Wiley-Blackwell, 2002).

33. Lee, W., von Hippel, P. H. & Marcus, A. H. Internally labeled Cy3/Cy5 DNA constructs show greatly enhanced photo-stability in single-molecule FRET experiments. *Nucleic Acids Res.* **42**, 5967–5977 (2014).

34. Heussman, D. *et al.* Measuring local conformations and conformational disorder of (Cy3)2 dimer labeled DNA fork junctions using absorbance, circular dichroism and two-dimensional fluorescence spectroscopy. *Faraday Discuss.* **216**, 211–235 (2019).

35. Cooper, G. M. *The Cell*. (Sinauer Associates, 2000).

36. Carlsson, N., Jonsson, F., Wilhelmsson, L. M., Nordén, B. & Åkerman, B. DNA hosted and aligned in aqueous interstitia of a lamellar liquid crystal – a membrane–biomacromolecule interaction model system. *Soft Matter* **9**, 7951–7959 (2013).

37. Agbavwe, C. & Somoza, M. M. Sequence-Dependent Fluorescence of Cyanine Dyes on Microarrays. *PLOS ONE* **6**, e22177 (2011).

38. Kretschy, N., Sack, M. & Somoza, M. M. Sequence-Dependent Fluorescence of Cy3- and Cy5-Labeled Double-Stranded DNA. *Bioconjug. Chem.* **27**, 840–848 (2016).

39. Laptenok, S. P. *et al.* Global analysis of Förster resonance energy transfer in live cells measured by fluorescence lifetime imaging microscopy exploiting the rise time of acceptor fluorescence. *Phys. Chem. Chem. Phys.* **12**, 7593–7602 (2010).

40. Lambert, T. J. FPbase: a community-editable fluorescent protein database. *Nat. Methods* **16**, 277–278 (2019).

41. Iqbal, A. *et al.* Orientation dependence in fluorescent energy transfer between Cy3 and Cy5 terminally attached to double-stranded nucleic acids. *Proc. Natl. Acad. Sci.* **105**, 11176–11181 (2008).

42. Hagerman, P. J. Flexibility of DNA. *Annu. Rev. Biophys. Biophys. Chem.* **17**, 265–286 (1988).

43. Allison, S. A. & Schurr, M. J. Torsion Dynamics and Depolarization of Fluorescence of Linear Macromolecules I. Theory and Application to DNAt. *Chem. Phys.* **41**, 35–59 (1979).

44. Lee, H. H. *et al.* Crystal structure of human nucleophosmin-core reveals plasticity of the pentamer–pentamer interface. *Proteins Struct. Funct. Bioinforma.* **69**, 672–678 (2007).

45. Šašinková, M. *et al.* NSC348884 cytotoxicity is not mediated by inhibition of nucleophosmin oligomerization. *Sci. Rep.* **11**, 1084 (2021).

46. Kremers, G.-J., Goedhart, J., van Munster, E. B. & Gadella, T. W. J. Cyan and Yellow Super Fluorescent Proteins with Improved Brightness, Protein Folding, and FRET Förster Radius,. *Biochemistry* **45**, 6570–6580 (2006).

47. Heesink, G. *et al.* Quantification of Dark Protein Populations in Fluorescent Proteins by Two-Color Coincidence Detection and Nanophotonic Manipulation. *J. Phys. Chem. B* **126**, 7906–7915 (2022).

48. Brida, D., Manzoni, C. & Cerullo, G. Phase-locked pulses for two-dimensional spectroscopy by a birefringent delay line. *Opt. Lett.* **37**, 3027 (2012).

49.     Réhault, J., Maiuri, M., Oriana, A. & Cerullo, G. Two-dimensional electronic spectroscopy with birefringent wedges. *Rev. Sci. Instrum.* **85**, 123107 (2014).

50.     Mooney, J. & Kambhampati, P. Get the Basics Right: Jacobian Conversion of Wavelength and Energy Scales for Quantitative Analysis of Emission Spectra. *J. Phys. Chem. Lett.* **4**, 3316–3318 (2013).

51.     Brodská, B., Holoubek, A., Otevřelová, P. & Kuželová, K. Low-Dose Actinomycin-D Induces Redistribution of Wild-Type and Mutated Nucleophosmin Followed by Cell Death in Leukemic Cells. *J. Cell. Biochem.* **117**, 1319–1329 (2016).

52.     Holoubek, A. *et al.* AML-Related NPM Mutations Drive p53 Delocalization into the Cytoplasm with Possible Impact on p53-Dependent Stress Response. *Cancers* **13**, 3266 (2021).

**Acknowledgement**

We thank Adil Haboucha, Sebastian Claudote and David Méchin from Photonics Bretagne for providing the photonic crystal fiber for broadband white light generation. This project has received funding from the European Union's Horizon 2020 research and innovation programme under the Marie Skłodowska-Curie grant agreement No. 101030656. PH, DS and AH acknowledge the Czech Science Foundation grant No. 22-03875S.

**Methods**

*FRET-ixFLIM description*

In a typical FRET experiment, there is a pair of fluorescent molecules called donor and acceptor. After the excitation by the laser pulse, there will be a population of excited donors $P_D(0)$ and excited acceptors $P_A(0)$. If we assume rate kinetics, the equations for the following excited-state population dynamics are

$$\frac{dP_D(t)}{dt} = -k_D P_D(t) - k_T P_D(t)$$

$$\frac{dP_A(t)}{dt} = +k_T P_D(t) - k_A P_A(t).$$

Here, we have the rates $k_A, k_D$ for the decay of the isolated acceptor and donor, respectively, and the rate of the Förster transfer $k_T$. The solution is

$$P_D(t) = P_D(0) e^{-(k_D + k_T)t}$$

$$P_A(t) = P_A(0) e^{-k_A t} + P_D(0) \frac{k_T}{k_T + k_D - k_A} \left( e^{-k_A t} - e^{-(k_T + k_D)t} \right). \quad (7)$$

The initial populations are proportional to the excitation spectra of the acceptor and donor $\epsilon_A(\omega)$ and $\epsilon_D(\omega)$ and their concentrations in the sample $c_A$, $c_D$ (for brevity we drop the $\tau$ subscript by the frequency $\omega_\tau$ in this section). First, we assume that a long-pass filter and sufficient spectral separation allow detection of the acceptor emission only, $FL(t) \propto P_A(t)$. The transient excitation spectral map can then be expressed as

$$\text{ixFLIM}_A(\omega, t) = k_A \phi_A \left\{ c_A \epsilon_A(\omega) e^{-k_A t} + c_D \epsilon_D(\omega) \frac{k_T}{k_T + k_D - k_A} \left( e^{-k_A t} - e^{-(k_T + k_D)t} \right) \right\}. \quad (8)$$

where $k_A\phi_A$ expresses the number of photons detected from emitting population $P_A$ per time, with $\phi_A$ is the acceptor FL quantum yield multiplied by the total detection efficiency. This expression applies to our cyanine-dye-labelled beads, with $c_A = c_D$ by design (donor and acceptor on complementary strands of the double-stranded DNA). Clearly, in this case the mere presence of $\epsilon_D(\omega)$ in the spectrum implies $k_T \neq 0$, i.e., directly reports on transfer. The acceptor spectrum has in ixFLIM always simple kinetics, decaying with the acceptor lifetime. The donor spectrum, however, exhibits the competition of donor-acceptor transfer and acceptor decay. For $k_T + k_D \gg k_A$ (often called normal kinetics), the signal first rises with the rate of $k_T + k_D$ and then decays with the lifetime of the acceptor, i.e., the rate $k_A$. For $k_T + k_D \ll k_A$ (often called inverted kinetics), the transferred excitation rapidly decays without building up acceptor population. In this case, the donor spectrum will at longer times decay with the shortened lifetime of the donor, i.e., rate $k_T + k_D$. Unlike in transient absorption, where the inverted kinetics is hard to observe due to the small buildup of $P_A$, both cases of transfer lead to a strong signal. The normal kinetics is exemplified by the cyanine-dye-labelled beads, where all signals clearly decay with the lifetime of the acceptor at longer times. Upon integration over detection time, we obtain

$$\text{ixFLIM}_A(\omega) = \int_0^\infty ix\text{FLIM}(\omega, t)dt = \phi_A \left\{ c_A\epsilon_A(\omega) + c_D\epsilon_D(\omega) \frac{1}{1 + \frac{k_D}{k_T}} \right\}.$$

Here, we immediately recognize in the last fraction the transfer efficiency $E$ (Eq. (3) in the main text), getting

$$\text{ixFLIM}_A(\omega) = \phi_A \{ c_A\epsilon_A(\omega) + c_D\epsilon_D(\omega)E \}. \quad (9)$$

In case that appreciable ratio $\frac{\phi_D}{\phi_A}$ of the donor emission is detected as well, a term

$$\text{ixFLIM}_D(\omega, t) = \phi_D k_D c_D \epsilon_D(\omega) e^{-(k_D + k_T)t} \quad (10)$$

has to be added to the expression Eq. (2) for ixFLIM$_A(\omega, t)$ above, and the total signal is

$$\text{ixFLIM}(\omega, t) = \text{ixFLIM}_A(\omega, t) + \text{ixFLIM}_D(\omega, t). \quad (11)$$

The implication is that the rise component in the donor spectrum is decreased by a factor of $\frac{\phi_D k_D}{\phi_A k_A}$. This is straightforward to understand since this fraction leaks through the filter and it is detected whether transferred to the acceptor or not. In the time-integrated signal (Eq. 9), this translates to a correction to the extracted transfer efficiency $E$:

$$\text{ixFLIM}(\omega) = \phi_A \left\{ c_A\epsilon_A(\omega) + c_D\epsilon_D(\omega) \left[ E + \frac{\phi_D}{\phi_A}(1 - E) \right] \right\}.$$

In more complicated samples, such as our HEK-293T cells with nucleophosmin labeled with mVenus and mRFP1, the situation can be more complex. The main complication is that there are multiple fractions of the emitting populations. In most cases, the population fractions will be independent of each other, i.e., each one will obey its dynamics. ixFLIM has the advantage of easily isolating other species $s$ than the desired donor-acceptor pair, since these will contribute as $k_s \phi_s c_s \epsilon_s(\omega) e^{-k_s t}$ and can be identified by their spectrum, their lifetime of both. The species *s* can either be other molecules (e.g., autofluorescence of the cells), or a fraction of the donor/acceptor molecules with varying transfer efficiency and/or varying excited-state decay. For each pair of donor and acceptor, the dynamics in Eq. (1) can be solved, yielding two components (spectral and temporal) to the overall dynamics. While this can complicate the analysis, the spectral separation is still helpful. In the acceptor spectral region, only the acceptor decay components are present. These can be fit first and then fixed in the global fit. The spectra in the donor region then report on the interaction in case they exhibit rise (with the rate $k_T + k_D$) and/or decay with the lifetime of the acceptor obtained from the first fit. This is the case of our labelled HEK-293T cells, where the FRET presence indicates interaction and thus NPM oligomerization. The complete ixFLIM form for the labelled NPM with fractions of free donor and acceptor is

$$ixFLIM(\omega, t) = k_A \phi_A \left\{ c_A \epsilon_A(\omega) e^{-k_A t} + c_D^b \epsilon_D(\omega) \frac{k_T}{k_T + k_D - k_A} \left( e^{-k_A t} - e^{-(k_T + k_D)t} \right) + \frac{k_D \phi_D}{k_A \phi_A} \epsilon_D(\omega) \left( c_D^f e^{-k_D t} + c_D^b e^{-(k_T + k_D)t} \right) \right\}. \quad (12)$$

Here, we separated the concentration of the donor to the bound $c_D^b$ and free $c_D^f$. For acceptor only the total concentration matters. Clearly, the amplitude of the rise component with the donor spectrum is decreased by the part of bound donor emission. On the other hand, the component with the acceptor decay time is insensitive to the emission from the donor.

*Optical setup*

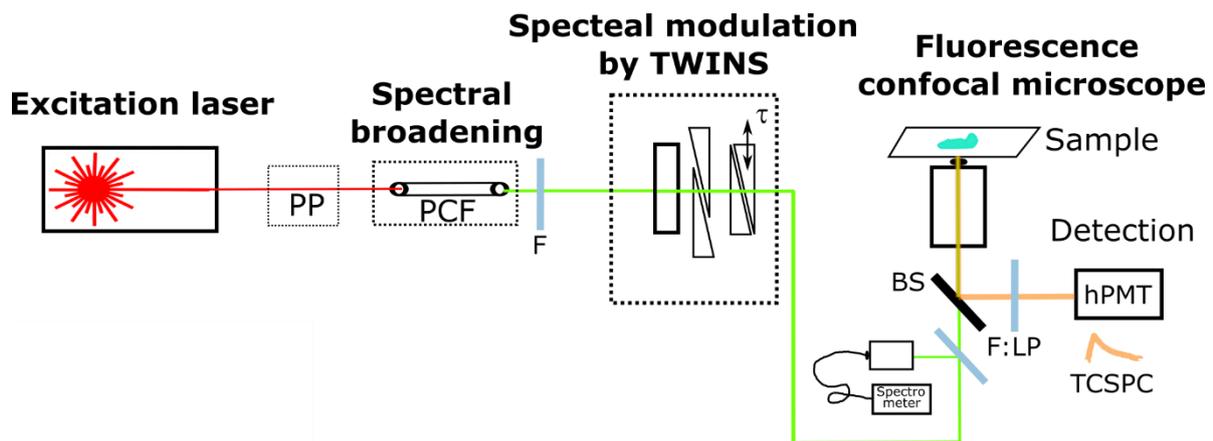

**Figure 6 | Experimental setup.** Excitation laser (Coherent Chameleon Ultra II) pulses (~150 fs, 760 nm, 4 MHz, 4 nJ after pulse picker (PP, PulseSelect from APE) and attenuation) seed the photonic crystal fiber (PCF, custom-made in collaboration with Photonics Bretagne), which generates spectrally broad (~450 nm to 700 nm after long- and short-pass filtering (F, Thorlabs)) spectrum. The broadband pulses enter the home-built TWINS interferometer used for creating a double-pulse with a variable inter-pulse delay $\tau$. Fraction of the pulses is split off and coupled by a 10x objective to a fibre spectrometer (SR-2, Ocean Insight), where the spectrum is used for reference and calibration. The pulses are further attenuated and coupled into a commercial confocal scanning fluorescence microscope (Olympus iX83 with FV1200 scanner), where they excite the sample by apochromatic objective (Olympus). The fluorescence emission is split off by a beam splitter, the residual excitation is removed by a long-pass filter (F:LP), and detected by hybrid photomultiplier (PMA hybrid 40, PicoQuant) counted by a TCSPC module (TimeHarp 260 PICO, PicoQuant).

Our experimental setup is schematically shown and described in Fig. 6. We use a custom-built excitation coupled to a commercial microscope with FLIM. To facilitate high-enough photon counts per second that allow fast image scanning and sensitive acquisition of weak fluorescence signals at low excitation pulse energies, FLIM measurements are typically done at MHz pulse repetition rates. To create spectrally ultrabroadband pulses at such repetition frequencies we utilize a photonic crystal fiber (PCF, custom-made in collaboration with Photonics Bretagne, see details in the Methods section. Pumped by a Ti:Sapphire oscillator (Chameleon Ultra II, Coherent), our PCF spectrum ranges from 400 nm to 1000 nm, with a typical repetition rate of 4 MHz. Note, that alternative fiber-based white light pulsed lasers with similar parameters are commercially available. The spectrum is restricted to the desired window by a shortpass filter at the long-wavelength edge of the absorption. The pulse pair for spectral modulation of the excitation is produced interferometrically. We use a custom-built version of the birefringent common-path interferometer called TWINS[48,49], that features excellent both phase stability and beam pointing stability. For the working of the TWINS interferometer we refer the reader to Refs. [48,49], briefly, the delay between the ordinary and extraordinary waves in the birefringent $\alpha$-BBO crystal is used to separate the pulse into two time-delayed replicas. Translation of a pair of wedges into the beam alters the thickness of the material in which the pulses travel with different refractive index, and thus changes the inter-pulse delay $\tau$. The pulse pairs are coupled into the commercial scanning confocal fluorescence microscope (iX83 with FV1200scanner, Olympus) where they are used to excite the sample, using an apochromatic objective (Olympus UPlanSApo 10x for the Oxo VI and UPlanSApo 40x for the Cy3–Cy5-labeled beads and HEK-293T cells). The fluorescence is detected through a confocal pinhole in a de-scanned port by a cooled hybrid photomultiplier tube (PMA Hybrid 40, PicoQuant), counted by a TCSPC module (TimeHarp, PicoQuant). The intensity of excitation was attenuated such that the data collection rate was kept well below 5% of the 4 MHz repetition rate to avoid pile-up in the photon counting. The acquisition was realized by SymphoTime software (PicoQuant), triggered by a scanning software of the microscope (Olympus). The microscope scan is triggered by a custom-made program in Matlab (Mathworks), that coordinates the experiment and controls the other instruments as well (TWINS, fiber spectrometer). The TWINS interferometer is home-built using $\alpha$-BBO wedges (FOCTek) of the same parameters as in Ref. [49]. The moving pair of wedges is mounted on a piezo stage (V-408, PI) that allows rapid and precise scanning of the delay time $\tau$.

*ixFLIM signal processing*

The ixFLIM dataset $\text{ixFLIM}(x, y, t, \tau)$ is acquired in form of a stack of $N_\tau$ FLIM frames ($N_\tau$ is a number of $\tau$ steps). The data 'hypercube' is exported from SymphoTime (PicoQuant) software and further processed by a custom-made Matlab (Mathworks) script. For each pulse delay $\tau$ (and thus FLIM frame), the modulated excitation laser spectrum $\text{Spec}(\lambda, \tau)$ is collected (see Fig. 6 for the setup). In the first step, the $\tau = 0$ point has to be found and the TWINS interferometer step size calibrated. This is done from the Fourier-transformed reference spectra sequence $\text{Spec}(\lambda, \tau) \to Spec(\lambda, \lambda_\tau)$ in an automatized way, as described in detail in the SI. Once the $\tau = 0$ position and TWINS calibration is known, the ixFLIM stack can be Fourier-transformed into the frequency domain: $\text{ixFLIM}_{\text{raw}}(x, y, t, \omega_\tau) = \text{Re} \int d\tau e^{i\omega_\tau \tau} \text{ixFLIM}_{\text{raw}}(x, y, t, \tau)$. In order to suppress the zero frequency, the mean of the dataset along $\tau$ is first subtracted. The Fourier-transformed dataset is then converted from frequency to wavelength[50], including the Jacobi transform $d\omega = \frac{2\pi c}{\lambda^2} d\lambda$ that multiplies the converted spectrum by $\frac{1}{\lambda^2}$. Finally, the resulting full dataset $\text{ixFLIM}_{\text{raw}}(x, y, t, \lambda_\tau)$ is divided by the excitation laser spectrum, obtaining the ixFLIM dataset $\text{ixFLIM}(x, y, t, \lambda_\tau) = \frac{\text{ixFLIM}_{\text{raw}}(x,y,t,\lambda_\tau)}{\text{Spec}(\lambda_\tau)}$. This is then used for imaging and further analysis, as described in the paper.

The ixFLIM dataset $\text{ixFLIM}(x, y, t, \omega_\tau)$ includes as its marginals standard FLIM: $\text{FLIM}(x, y, t) = \int d\omega_\tau \text{ixFLIM}(x, y, t, \omega_\tau)$, as well as excitation-spectrum imaging (interferometric excitation fluorescence imaging, ixFIM): $\text{FIM}(x, y, \omega_\tau) = \int dt \text{ixFLIM}(x, y, t, \omega_\tau)$.

Wherever mean lifetime is calculated, the first moment is meant:

$$\langle t \rangle = \frac{\int (t - t_0) \text{FLIM}(t) dt}{\int \text{FLIM}(t) dt},$$

where $t_0$ is the time when the excitation pulse arrives.

Analogously, the mean excitation wavelength is calculated as center of mass,

$$\langle \lambda_\tau \rangle = \frac{\int \lambda_\tau \text{FIM}(\lambda_\tau) d\lambda_\tau}{\int \text{FIM}(\lambda_\tau) d\lambda_\tau},$$

where the integration is over the wavelength region covered by the excitation laser spectrum.

*Cy3-Cy5 DNA Constructs*

A set of five 27nt long DNA oligonucleotides was designed. The set consists of two complementary DNA strands, which can hybridize to form double-stranded DNA.

5´-GTGAGT$^6$AAGAGATACACATGGATTGAG-3´ oligonucleotide was biotinylated on its 5´-end and internally labeled with Cy3 fluorescent dye (fluorescence donor) at a position 6 (as marked by superscript number in the sequence above), this strand is referred to as D6. Next, four oligonucleotides with identical nucleotide sequence but different Cy5 (fluorescent acceptor) position were prepared: 5´-CT$^2$CAATCCA$^9$T$^{10}$GTGTATCT$^{18}$CTTACTCAC-3´ internally labeled with Cy5 at positions 2 (A2), 9 (A9), 10 (A10) and 18 (A18), respectively (as marked by superscript numbers in the sequence above). All oligonucleotides were synthesized and labeled by Integrated DNA Technologies, Inc. When hybridized together at 60°C in GeneQ Thermal Cycler (Bioer Technology Co.) we obtained 4 dsDNA strands, which differed in fluorescent donor Cy3 and acceptor Cy5 distance: the distance was 4 bp for D6:A18, 12 bp for D6:A10, 13 bp for D6:A9 and 20 bp for D6:A2 dsDNA. Using biotinylated 5´-end of the D6 oligonucleotide, individual dsDNA strands were bound to Dynabeads® M-270 Streptavidin magnetic beads (Thermo Fisher Scientific) according to the standard protocol. After washing, labeled beads were resuspended in 50 mM Tris·HCl, pH 7.5, to the desired concentration.

*Cell cultivation, transfection and fixation*

Adherent HEK-293T cells were cultured at 37 °C under standard cultivation conditions in RPMI growth medium (Sigma) supplemented with 10% FBS (Biochrom) under 5% CO2 atmosphere. Plasmids for expression of NPMwt fused with mRFP1[51,52] and with mVenus[11] were amplified in E. coli and purified with PureYield Plasmid Miniprep System (Promega). The cells grown on a glass-bottom Petri dish were transfected using jetPrime transfection reagent (Polyplus Transfection) according to the manufacturer's protocol. The transfected cells were further grown for 24 h prior to cell fixation. The transfected cells were fixed with 4% paraformaldehyde and permeabilized by 0.5% Triton X-100 as described in Ref. [2]. After final wash, the cells were kept in the sterile PBS in 4 °C. Experiments with the fixed cells were done at room temperature.

**Supporting Information to:**

**ixFLIM: Interferometric Excitation Fluorescence Lifetime Imaging Microscopy**

Pavel Malý[1*], Dita Strachotová[1], Aleš Holoubek[2], Petr Heřman[1]

[1]Faculty of Mathematics and Physics, Institute of Physics, Charles University, Prague, Czech Republic

[2]Department of Proteomics, Institute of Hematology and Blood Transfusion, Prague, Czech Republic

*pavel.maly@mff.cuni.cz

**Ultrabroadband interferometry**

The broadband white light continuum is generated in a photonic crystal fiber (PCF)[1], custom-made in collaboration with Photonics Bretagne. The fiber is about 12 cm long, and, to prevent thermal damage to the fiber, we use end-caps that seal the ends of the PCF, preventing contamination and, at the same time, protecting against the surface burn since the focus lies on the end-cap–core interface and is not exposed to air. The PCF is pumped by about 4 nJ pulses, centered at 760 nm, at repetition rate of 4 MHz, producing spectrum typically ranging from 450 nm to 1000 nm. The spectrum is restricted by long- and short-pass filters to the desired spectral range.

Here, we describe the calibration procedure for the excitation by our home-built TWINS[2] interferometer based on birefringent wedges with variable insertion into the beam. In the TWINS interferometer, the delay between the two pulses is scanned by a stage translating a pair of birefringent wedges[2,3]. For each step of the stage, the modulated spectrum is recorded just before entering the microscope, using a back-reflection from an ND filter and 10x objective coupling into fiber spectrometer. The resulting spectral interference is depicted in Fig. S1 left.

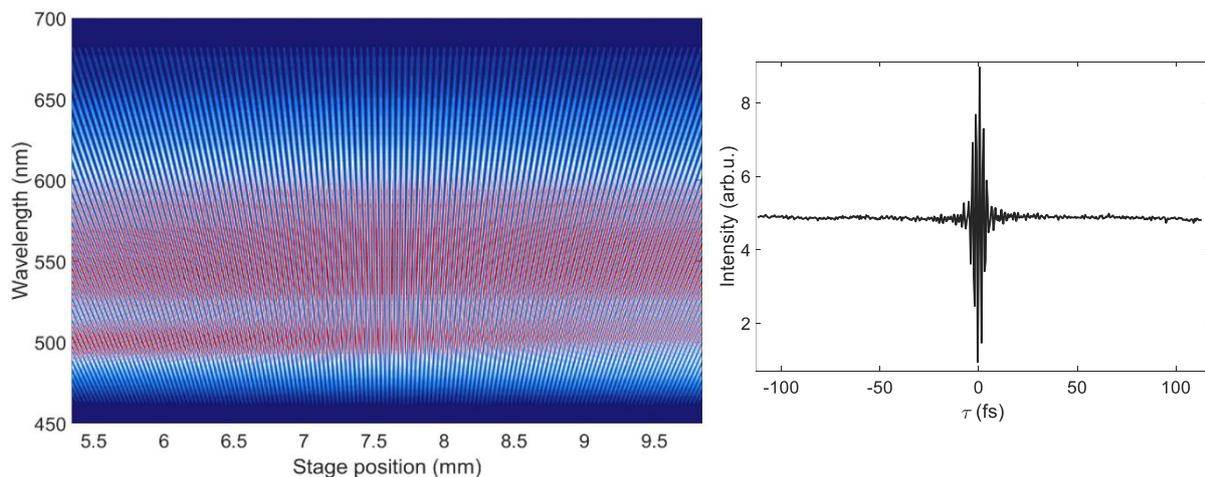

*Figure S1 Spectral interferometry for TWINS calibration and reference spectrum. Left: spectral interference of the pulse pair generated by the interferometer, as a function of the TWINS stage position. Right: Spectrally integrated interferogram of the same scan, already with axis translated into time delay $\tau$.*

Clearly, the tilt of the fringes is symmetric around a specific position of the stage just above 7.5 mm, which corresponds to the $\tau = 0$ point. To determine this point more precisely, we integrate the spectrum over the detection wavelength, obtaining the interferogram (Fig. S1 right). Fourier-transform of this interferogram has to yield the laser spectrum. We thus guess the $\tau = 0$ point from the maximum of the interferogram and iteratively adjust it, taking for each point the real part of the Fourier transform. The true $\tau = 0$ point will be the one for which is the spectrum without modulation resulting from delay phase twists and has maximum area under its curve. Having thus found the $\tau = 0$ point, we guess the stage step size as 50 fs/mm, getting the time delay axis used for the interferogram in Fig. S1 right. The scan step size and length are given by the Nyquist limit, determined by the lowest wavelength in the spectrum and the desired spectral resolution. In our measurement, we scan the interferogram symmetrically around $\tau = 0$ with stage steps of 0.015 mm (0.75 fs), measuring 301 steps (from -112.5 fs to 112.5 fs).

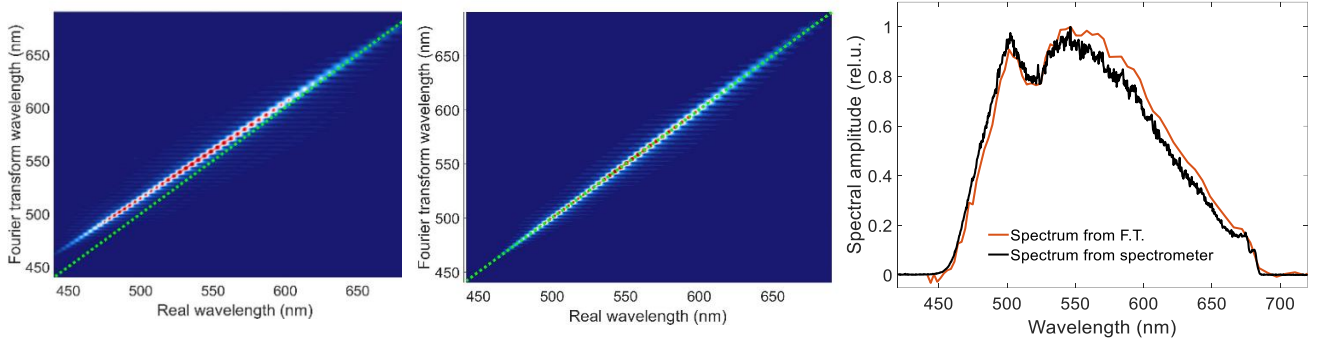

*Figure S2 Calibration of the interferometr. Left: uncalibrated Fourier-transformed spectra stack, middle: the same but with the F.T. wavelength axis calibrated as described in the text. The dotted green line indicates the desired wavelength correspondence. Right: Recovered spectrum from the interferogram in Fig. S1 right (orange), compared to the spectrometer-acquired spectrum (black).*

To obtain the spectral calibration, we and plot the spectral interference in Fourier domain correlating $\lambda_{FT}$ and $\lambda_{real}$ from the spectrometer (Fig. S2 left). Like this, we get one wavelength point (around 650 nm) for which the calibration is correct (for which the step size is indeed 0.75 fs). For the other wavelengths the stage step size is bit different due to dispersion of the wedge crystal. While one can account for the dispersion using the Sellmaier expansion of the refractive index[4], we choose to simply take the maximum point for each wavelength and fit the $\lambda_{FT}(\lambda_{real})$ by a polynomial (in practice linear dependence suffices). Inverting the fit, we get the calibration $\lambda_{real}(\lambda_{FT})$ that we use for the Fourier-transformed ixFLIM dataset. Comparing the spectrum from the Fourier-transformed interferogram (including the $\frac{1}{\lambda^2}$ factor[5]) from Fig. S1 right with the spectrum from the spectrometer (Fig. S2 right), the calibration can be checked.

Next to the calibration, the laser spectrum is used to calculate the ixFLIM data from the raw measured signal as as $\text{ixFLIM}(x, y, t, \lambda_\tau) = \frac{\text{ixFLIM}_{\text{raw}}(x,y,t,\lambda_\tau)}{\text{spec}(\lambda)}$.

## Information in ixFLIM: Oxonol VI binding to albumin

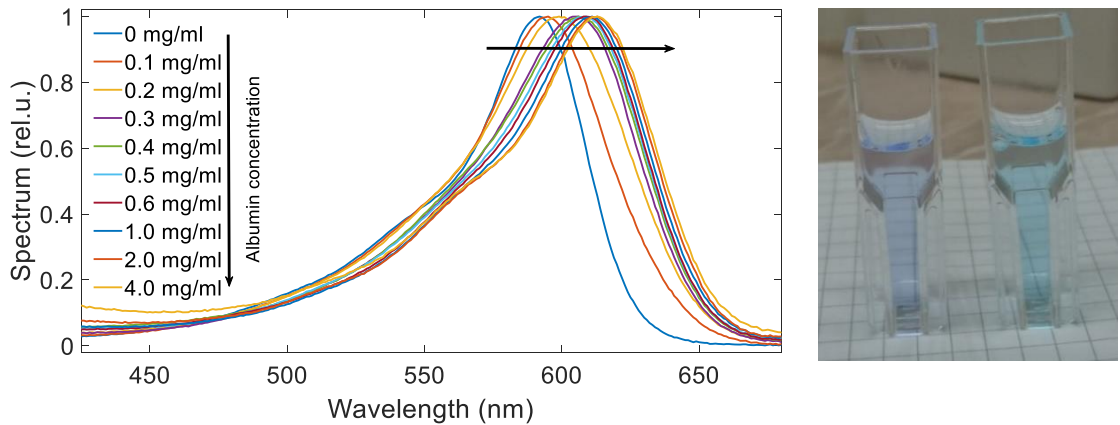

*Figure S3 Shift of the Oxonol VI absorption spectrum with bovine serum albumin (BSA) concentration. Left: Titration study. Clearly, above 2 mg/ml the solution is saturated, and all Oxo VI is bound to BSA. Right: The solutions with free (left, purple) and bound (right turquoise) Oxo VI.*

The absorption spectra of Oxo VI were measured on absorption spectrometer Varian Cary50 UV/VIS spectrometer.

## FRET-ixFLIM molecular ruler: Cy3-Cy5 at DNA

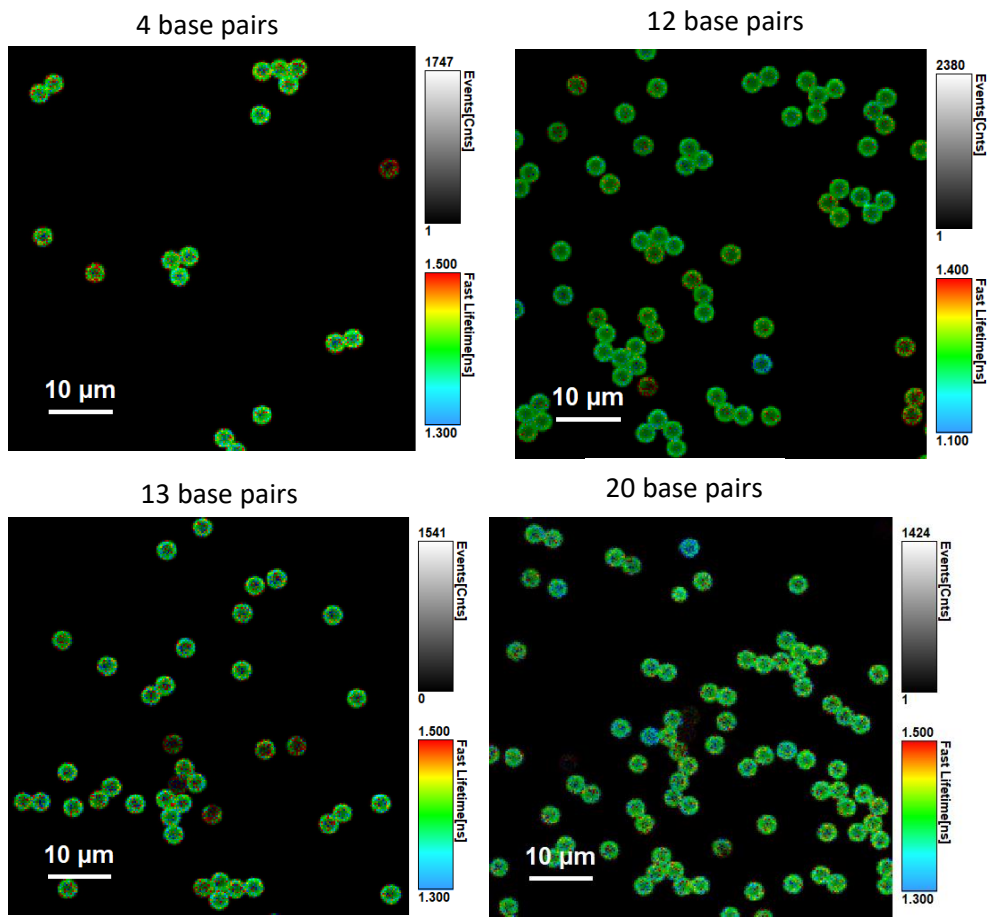

*Figure S4 FLIM images of the beads labelled with the DNA constructs. Clockwise: Cy3–Cy5 distance of 4 bp, 12 bp, 13 bp and 20 bp. Note the shorter lifetime for the 12 base pair beads. These beads were used for the ixFLIM analysis described here and in the main text.*

Four types of beads with different bound DNA constructs have been measured, with the Cy3 and Cy5 within distance of 4 bp, 12 bp, 13 bp and 20 bp, Fig. S4. While the donor Cy3 position remains fixed, the acceptor position Cy5 varies with the increasing distance. Interestingly, for the 12 base pair position, the Cy5 lifetime is shorter (1.2 ns instead of 1.4 ns), as reflected also by the global analysis (fits below). This can be explained by the variability of the Cy5 quantum yield (and lifetime) dependent on the basis-dependent immediate environment within the DNA[6,7].

For the spectral fits, the spectra were taken from the FPbase database[8] and slightly adjusted multiplying by skewed Gaussians to fit the 4 base-pair (13.6 Å) distant donor-acceptor pairs with favorable orientation factor, for which the efficiency is bound to be close to unity: the Förster critical radius is 53 Å, i.e., randomly-oriented dyes would have efficiency of 0.9997, and favorably oriented ones even higher. The spectra were used for the fitting of the FRET efficiency as described in the Methods section of the main text (Eq. (9) with $c_A = c_D$). To avoid the increase of noise close to the edges of the excitation spectrum by the spectral division, we did not divide the acquired excitation spectra by the laser spectrum, but multiplied the Cy3 and Cy5 absorption spectra instead. The resulting fits for 4 bp, 12 bp, 13 bp and 20 bp distant donor-acceptor pairs, integrated over the beads in Fig. S4, are shown in Fig. S5. The extracted efficiencies are given in Table 1 and Fig. 3h in the main text. The absorption spectra used, Fig. 4c in the main text, not only work for the fitting, but are in a good agreement with those measured by Lee et al.[9] (Fig. S5 right)

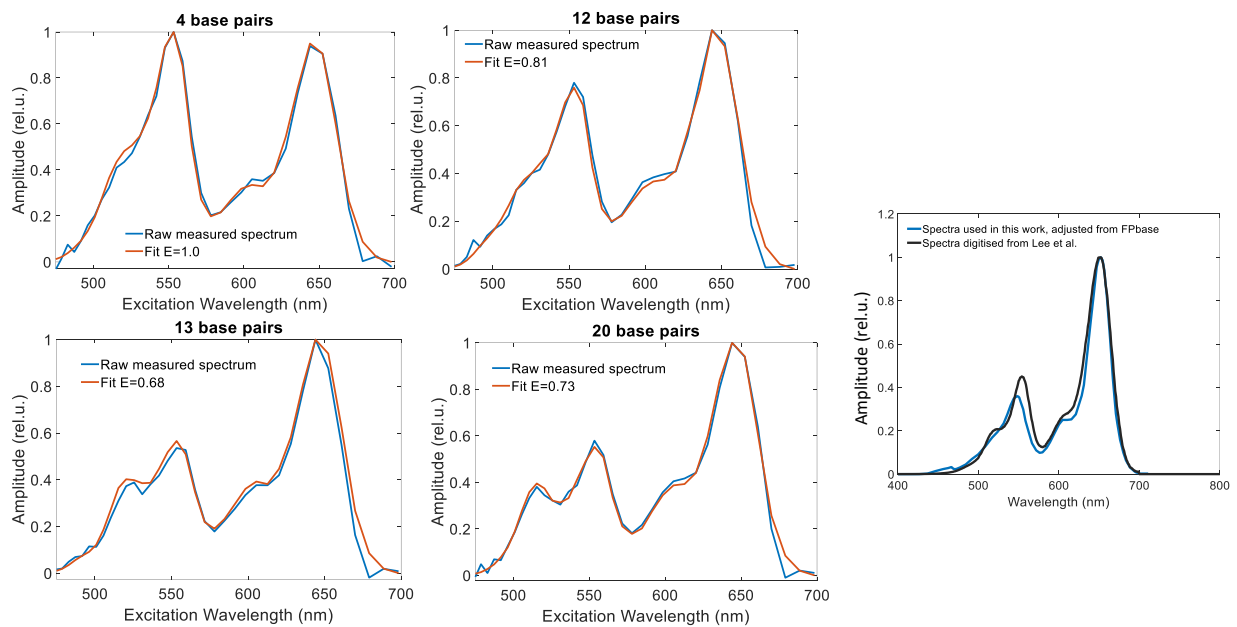

*Figure S5 Decomposition of the excitation spectra into those of the donor and acceptor. Left: the fits of the raw (not divided by the laser spectrum) $ixFIM(\omega_\tau)$ spectra by the (laser-spectrum multiplied) sum of the donor and acceptor absoprtion spectra, with the donor weighted by E. Right: the absorption (not excitation) spectrum of the Cy3 and Cy5 pair as reconstructed from the sum of the donor and acceptor spectra used in this work (blue), compared to the spectrum measured by Lee et al. (Ref. 9, black).*

For all the DNA constructs, the whole spatially integrated maps $\text{ixFLIM}(t, \omega_\tau)$ are shown in Fig. S6. These were fit by global analysis using two time components, see Fig. S7. These correspond to the donor excitation signal rise due to FRET (black) and acceptor emission decay (red). Note the shorter acceptor lifetime for the 12 bp distant dyes, due to the different position of the Cy5 acceptor within the DNA helix.

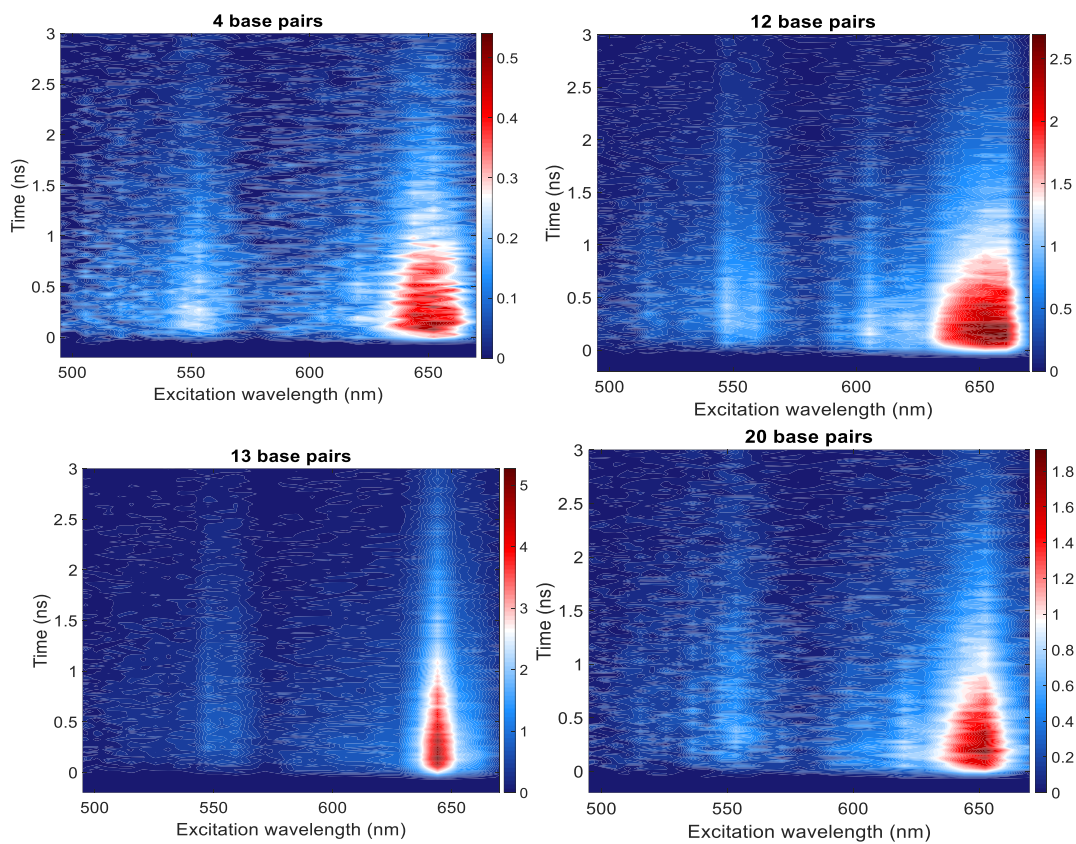

*Figure S6 Transient $ixFLIM(t, \lambda_\tau)$ maps for the four Cy3–Cy5 labelled DNA constructs imaged in Fig. S4.*

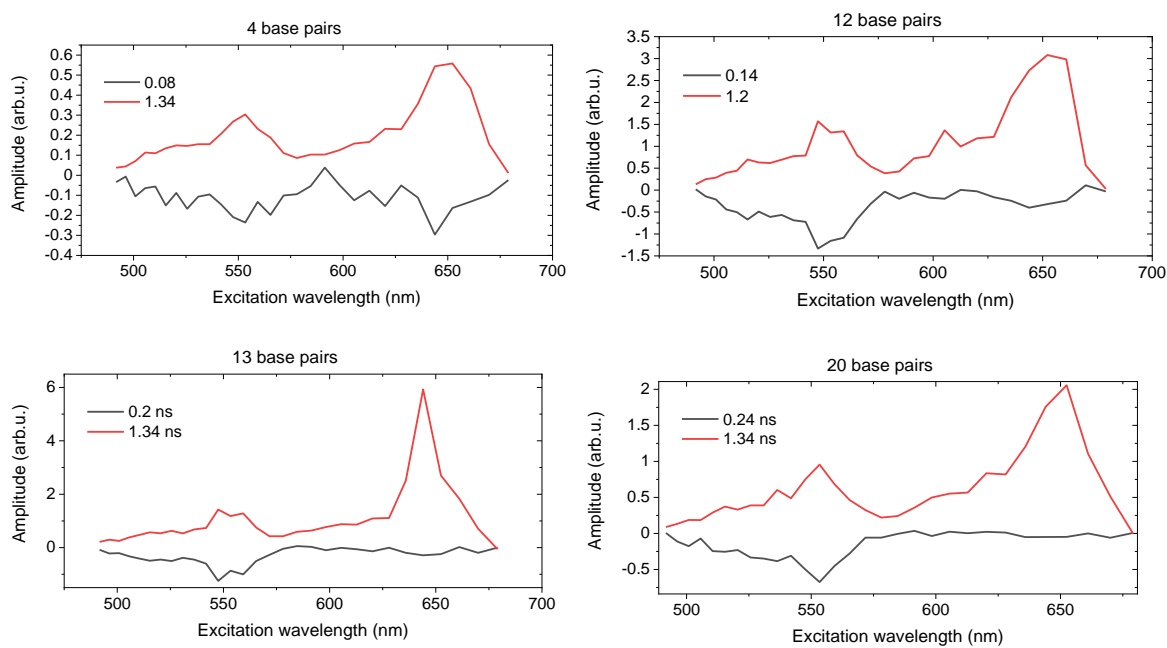

*Figure S7 Decay-associated spectra of the two time components from the global analysis of the transient maps shown in Fig. S6.*

The lifetime of Cyanine dyes is known to change when bound to DNA, dependent on the specific attachment site[6,7]. While the Cy5 acceptor position changes across the constructs, the Cy3 donor attachment remains the same, and thus its lifetime should remain the same as well. We measured the lifetime independently, on beads labeled by the single-stranded DNA with the Cy3 donor only (Fig. S8 left). The fluorescence decay of the non-transferring Cy3 donor can be fit with at minimum two exponentials, getting a mean lifetime of $\langle\tau\rangle = 0.82$ ns. This value is used for calculation of the transfer time from the rise of the signal from the interacting donor.

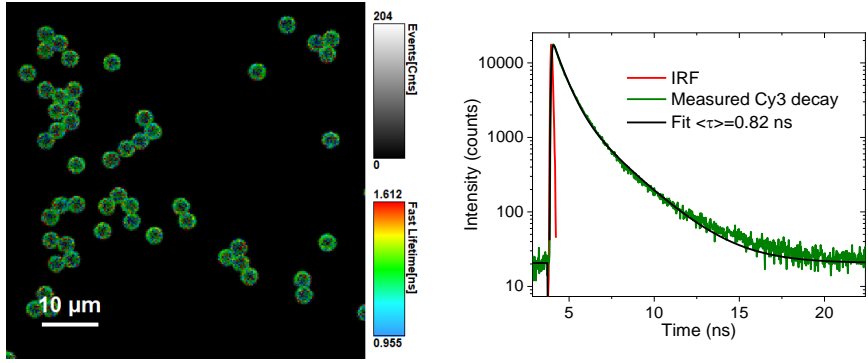

*Figure S8 Beads colored by Cy3 only (left), and their global fluorescence decay (right). The measured decay (green) was fit with two exponentials (black), reconvoluted with the separately-measured instrument response function (red). The slight long-time misfit reflects the absence of the weak background not originating from the Cy3 but from weak emission from the beads themselves, as we verified by a bead-only control measurement (data not shown).*

In the theoretical FRET model, we take the Cy3 donor and Cy5 acceptor dyes as represented by their transition dipole moments, attached to the DNA double helix on oppposite strands. We take the parameters of the B-type DNA: 10.5 pairs per helix turn, 0.34 helical rise per base pair, 10 Å diameter[10]. We consider the dyes to be attached as in Fig. 3a of the main text[9], with a helical rise angle $\alpha$, azimuthal angle $\phi$, and additional distance of the molecule from the helix radius $\Delta$. We further consider take the Förster critical radius to be $R_0 = 53$ Å and calculate the Förster rate using Formulas (1) and (2) in the main text. The important variables are the inter-dye distance $R$ and their mutual angle, given by $\alpha_{Cy3,Cy5}$ and $\theta_{Cy3,Cy5}$. Due to geometrical fluctuations and the fixed attachment site, the Cy3 and Cy5 angles $\alpha, \phi$ are taken to be disordered with a Gaussian distribution of angles around their central values given by the fixed helix geometry. For each realization of the angle disorder (taken to be the same for both of the dyes), we calculate the orientation factor $\kappa^2 = \hat{\mu}_a \hat{\mu}_d - 3(\hat{\mu}_a \cdot \hat{R})(\hat{\mu}_d \cdot \hat{R})$ and the inter-dye distance $R$, and from these the rate scaled to the known $k_T(R_0) = k_D$ from the definition of the $R_0$ as the distance where the FRET efficiency $E$ drops to half, $k_T = k_D \kappa_0^{-2} \left(\frac{R}{R_0}\right)^6$. For each realization of the disorder, we use this rate to calculate the transfer efficiency $E = \frac{k_T}{k_T + k_D}$, and we also calculate the time-dependent decay of the rising transfer component $S = \frac{k_T}{k_T + k_D - k_A} e^{-(k_T + k_D)t}$. The quantities $E$ and $S$ are then averaged over the normal-distributed angle disorder to produce the average efficiency $\langle E \rangle$ and average signal rise component $\langle 1 - \frac{k_T}{k_T+k_D-k_A} e^{-(k_T+k_D)t} \rangle$. The average efficiency is directly compared to the experimental data, the signal rise is, same as in the experiment, fit with an exponential to extract the $k_T$ and from this the efficiency. The values extracted in these two ways are compared to the experimental values, calculating the total mean-square deviation MSD. We globally minimize this MSD($\delta\alpha, \delta\phi, \Delta$) as a function of the disorder in the helical rise angle $\delta\alpha$, disorder of the azimuthal angle $\delta\phi$, and the additional distance of the dye transition dipole from the helix radius $\Delta$. The optimal values are $\delta\alpha = 0$, $\delta\phi = 27°$, and $\Delta = 0.6$ Å. The MSD cuts are shown in Fig. S9 for $\delta\alpha = 0$ (Fig. S9 left) and for $\Delta = 0.6$ Å (Fig. S9 right) with the lowest values indicated.

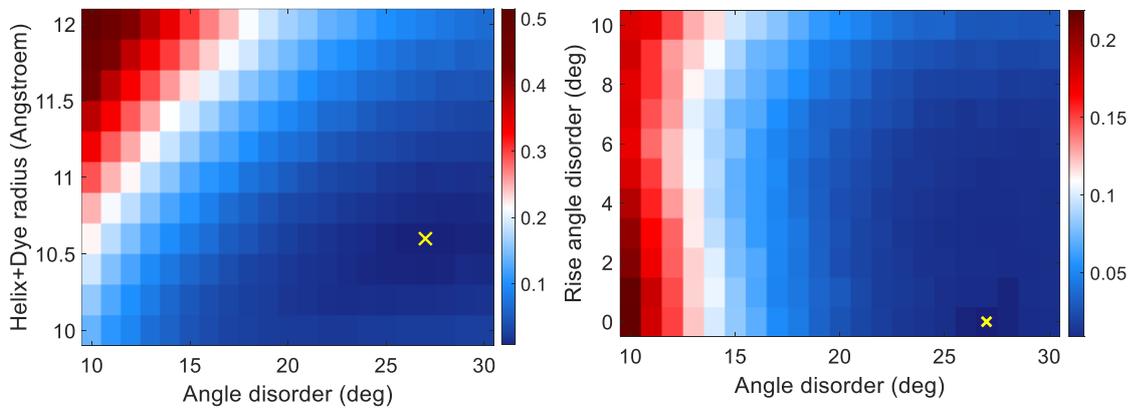

*Figure S9 Optimization of the parameters of the Cy3–Cy5 FRET calculation. Mean square deviation as a function of the fit values, cuts for fixed $\delta\alpha = 0$ (left) and for $\Delta = 0.6$ Å (right). Yellow crosses indicate the optimal lowest-MSD points.*

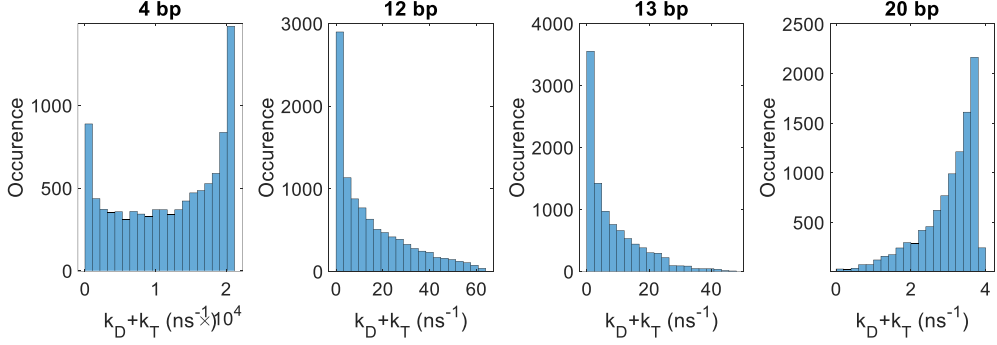

*Figure S10 Calculated distributions of the donor decay rate for the four donor-acceptor positions. Note the totally non-Gaussian character of the distributions arising from the normal distribution of the dye angles.*

Based on the central limit theorem, the distribution of the donor and acceptor angles is typically considered to be normally-distributed, as we do in our model. Interestingly, since there are only a few parameters that the FRET rate depends on, the Gaussian angle distribution leads to a highly non-Gaussian distribution of the transfer rates (see Fig. S10). For such distributions, it matters precisely what quantity dependent on the rate is averaged. In the ixFLIM experiment, the total signal with the donor excitation spectrum is described (as derived in the methods section):

$$ixFLIM(\omega, t) = k_A \phi_A c_D \epsilon_D(\omega) \sum_{i=1}^{N_{ens}} \left\{ \frac{k_T^i}{k_T^i + k_D - k_A} \left( e^{-k_A t} - e^{-(k_T^i+k_D)t} \right) + \frac{\phi_D k_D}{\phi_A k_A} e^{-(k_T^i+k_D)t} \right\}.$$

Here, we included the sum over all present transfer rates $k_T^i$. We have access to two measured quantities. First, we fit the time-integrated signal in the spectral domain, getting

$$ixFLIM\_Efit = \phi_A c_D \epsilon_D(\omega) \left\{ \left[ \langle E \rangle + \frac{\phi_D}{\phi_A} (1 - \langle E \rangle) \right] \right\},$$

i.e., the efficiency is averaged. The spectral-leak term (second in the square bracket) is very small here ($\frac{\phi_D}{\phi_A} = 0.002$) and can thus be safely neglected:

$$ixFLIM_{Efit} = \phi_A c_D \epsilon_D(\omega) \langle E \rangle.$$

Second, we fit (in the global analysis) the time kinetics and focus on the rising component:

$$ixFLIM_{tfit} = -k_A \phi_A c_D \epsilon_D(\omega) \langle \left[ \frac{k_T^i}{k_T^i + k_D - k_A} - \frac{\phi_D k_D}{\phi_A k_A} \right] e^{-(k_T^i+k_D)t} \rangle.$$

Again, without the spectral leak:

$$ixFLIM_{tfit} = -k_A \phi_A c_D \epsilon_D(\omega) \langle \frac{k_T^i}{k_T^i + k_D - k_A} e^{-(k_T^i + k_D)t} \rangle.$$

What gets averaged are thus the individual exponential rise components, weighted by a factor dependent on the transfer rate. In effect, fast rises get weight close to one, while the slow ones contribute much less to the total signal. The time-fitted spectra thus effectively produce larger apparent transfer efficiency, as is seen from Fig. 3h in the main text.

**Protein interaction: Nucleophosmin in nuclei and nucleoli of live HEK-293T cells**

As described in the main text, the total fluorescence decay of the representative HEK-293T cell nucleus can be fit with three exponential decay components, Fig. S11.

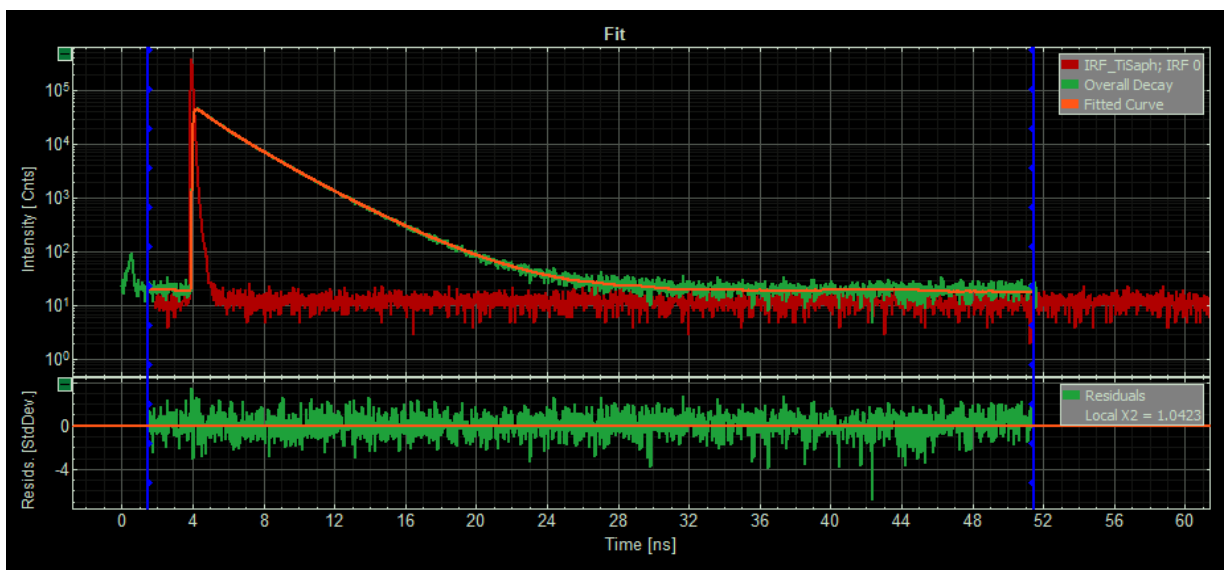

*Figure S11 Overall fluorescence decay (green) of the HEK-293T cell nucleus analyzed in the main text. The fit (orange) is using three exponentials with time constants of 0.7 ns, 1.81 ns and 2.94 ns, reconvoluted with the separately measured instrument response function (red). The fit and plotting as done in program SymphoTime (PicoQuant).*

These time constants were used for the global analysis of the transient $ixFLIM(t, \omega_\tau)$ maps (Fig. S12 top). The maps were analysed by global analysis for the whole nucleus, for the nucleoli only, and for the nucleoplasm only. The decay-associated spectra of the three time components are shown in Fig. S12 bottom.

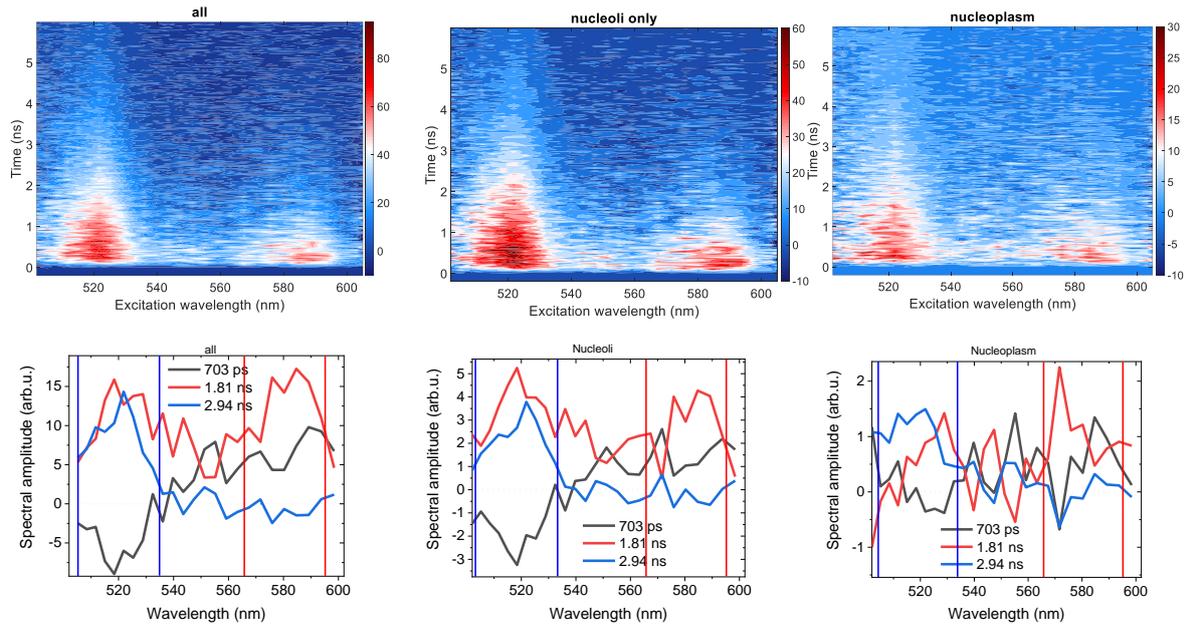

*Figure S12 ixFLIM of the dual-labelled NPM. Top: Global transient $ixFLIM(t, \omega_\tau)$ maps from the whole nucleus imaged in Fig. 5 in the main text (left), from the nucleoli only (middle) and from the surrounding nucleoplasm (right). Bottom: decay-associated spectra of the three time components from the global analysis of the transient maps, with the indicated time constants. Regions of the donor (blue) and acceptor (red) spectra integrated for the analysis are indicated by vertical lines.*

The intensity of the time component spectra in the donor (505 – 535 nm) and acceptor (565 – 595 nm) regions is summarized in the tables below and was used to infer the fraction of interacting mVenus donor as described in the main text. We found that, in the studied nucleus, the fraction of the interacting donor is $f_D^{\text{int}}|_{\text{nucleoplasm}} = 0.21$ in the nucleoplasm and $f_D^{\text{int}}|_{\text{nucleoli}} = 0.44$ in the nucleoli. This is expected from the significantly larger NPM concentration in the nucleoli, where it localizes.

| **all** | 0.7 ns | 1.81 ns | 2.94 ns |
|---|---|---|---|
| 505 – 535 nm | -1.89E-04 | 4.31E-04 | 2.88E-04 |
| 565 – 595 nm | 1.22E-04 | 3.11E-04 | -3.39E-05 |

| **nucleoli** | 0.7 ns | 1.81 ns | 2.94 ns |
|---|---|---|---|
| 505 – 535 nm | -1.47E-04 | 3.54E-04 | 1.93E-04 |
| 565 – 595 nm | 1.13E-04 | 2.28E-04 | -1.95E-05 |

| **nucleoplasm** | 0.7 ns | 1.81 ns | 2.94 ns |
|---|---|---|---|
| 505 – 535 nm | 9.11E-06 | 5.20E-05 | 9.52E-05 |
| 565 – 595 nm | 4.42E-05 | 7.52E-05 | -6.29E-08 |

**SI References**

1. Dudley, J. M., Genty, G. & Coen, S. Supercontinuum generation in photonic crystal fiber. *Rev. Mod. Phys.* **78**, 1135 (2006).

2. Brida, D., Manzoni, C. & Cerullo, G. Phase-locked pulses for two-dimensional spectroscopy by a birefringent delay line. *Opt. Lett.* **37**, 3027 (2012).

3. Réhault, J., Maiuri, M., Oriana, A. & Cerullo, G. Two-dimensional electronic spectroscopy with birefringent wedges. *Rev. Sci. Instrum.* **85**, 123107 (2014).

4. Thyrhaug, E. *et al.* Single-molecule excitation–emission spectroscopy. *Proc. Natl. Acad. Sci.* **116**, 4064–4069 (2019).

5. Mooney, J. & Kambhampati, P. Get the Basics Right: Jacobian Conversion of Wavelength and Energy Scales for Quantitative Analysis of Emission Spectra. *J. Phys. Chem. Lett.* **4**, 3316–3318 (2013).

6. Agbavwe, C. & Somoza, M. M. Sequence-Dependent Fluorescence of Cyanine Dyes on Microarrays. *PLOS ONE* **6**, e22177 (2011).

7. Kretschy, N., Sack, M. & Somoza, M. M. Sequence-Dependent Fluorescence of Cy3- and Cy5-Labeled Double-Stranded DNA. *Bioconjug. Chem.* **27**, 840–848 (2016).

8. Lambert, T. J. FPbase: a community-editable fluorescent protein database. *Nat. Methods* **16**, 277–278 (2019).

9. Lee, W., von Hippel, P. H. & Marcus, A. H. Internally labeled Cy3/Cy5 DNA constructs show greatly enhanced photo-stability in single-molecule FRET experiments. *Nucleic Acids Res.* **42**, 5967–5977 (2014).

10. DNA. *Wikipedia* https://en.wikipedia.org/wiki/DNA (2023).